\documentclass[%
 aip,
 amsmath,amssymb,
 reprint,%
author-year,%
]{revtex4-1}

\usepackage{graphicx}
\usepackage{dcolumn}
\usepackage{bm}

\usepackage[utf8]{inputenc}
\usepackage[T1]{fontenc}
\usepackage{mathptmx}
\usepackage{amsmath}
\usepackage{etoolbox}

\usepackage{color}
\usepackage{soul}
\usepackage{xfrac}

\makeatletter
\def\@email#1#2{%
 \endgroup
 \patchcmd{\titleblock@produce}
  {\frontmatter@RRAPformat}
  {\frontmatter@RRAPformat{\produce@RRAP{*#1\href{mailto:#2}{#2}}}\frontmatter@RRAPformat}
  {}{}
}%
\makeatother
\begin{document}
\renewcommand{\arraystretch}{2.2}

\preprint{AIP/123-QED}

\title[Similarity scaling of the axisymmetric turbulent jet]{Similarity scaling of the axisymmetric turbulent jet}
\author{Chunyue Zhu}
    \affiliation{College of Water Resources and Environmental Engineering, Zhejiang University of Water Resources and Electric Power, Hangzhou, China.}

\author{Yujia Tian}
    \affiliation{School of Mechanical and Aerospace Engineering,  Nanyang Technological University, Singapore 639798, Singapore.}

\author{Preben Buchhave}
    \affiliation{Intarsia Optics, S{\o}nderskovvej 3, 3460 Birkerød, Denmark.}

\author{Clara M. Velte}%
    \email{cmve@dtu.dk}
\affiliation{Department of Civil and Mechanical Engineering, Technical University of Denmark, Nils Koppels Allé, Building 403, 2800 Kongens Lyngby, Denmark.
}%

\date{\today}

\begin{abstract}
    In the current work, we find that a free axisymmetric jet in air displays self-similarity in the fully developed part of the jet. We report accurate measurements of first, second and third order, spatially averaged statistical functions of the axial velocity component performed with a laser Doppler anemometer, including in the outer (high intensity) regions of the jet. The measurements are compared to predictions derived from a simple jet model, described in a separate publication, and we discuss the implications for the further study of self-similarity in a free jet. It appears that all statistical functions included in this study can be scaled with a single geometrical scaling factor -- the downstream distance from a common virtual origin.
\end{abstract}

\maketitle

\section{Introduction}
Self-similarity, or self-preservation, is an important concept in the description of turbulent flows. Self-similarity can occur in turbulence when the flow is allowed to develop free of external influences so that the statistical properties are mainly determined by the local scales of the flow. Properties downstream can then, in those instances, be related to upstream properties by simple scaling factors. This significantly facilitates mathematical models and descriptions of such flows. It is also interesting from a more fundamental aspect, not least in terms of understanding the origins and underlying processes of self-similarity. In this paper we shall consider these questions with the turbulent, axisymmetric jet in air as a concrete and important example.

In a companion paper,~\cite{JetSimPart1}, we propose a simple model for an axisymmetric jet, free of external influences and issuing into quiescent air, and investigate what can be concluded about its statistical properties based on fundamental symmetry properties of space and time. Based on this analysis, we present a list of expected scaling parameters for some first, second and third order statistical functions, see Table~\ref{tab:1} in section~\ref{sec:TheJetModel}. We then describe the laboratory jet experiment and the laser Doppler anemometer (LDA) as well as the signal- and data-processing. The LDA is developed to be highly suited for velocity measurement in a flow with high turbulence intensity. We present results of measurements in a high Reynolds number jet in air and throughout the paper, we compare the measured results to the properties expected from the simple model, and finally present some general conclusions.

That turbulence has a tendency in some situations to develop into a state, which to some degree shows self-similarity for all or some of the statistical quantities describing the flow, has been known since the early pioneering theories for turbulence as a stochastic process, c.f.~\cite{Prandtl1925,von1930mechanische,KarmanHowarth,taylor1939some}. Among the first to define the concept of self-similarity in general, we can also mention~\cite{zel1937limiting}. The free jet has been studied as a canonical flow, well suited for laboratory experiments, and used as a benchmark for the study of self-similarity. Early studies of the axial jet were performed by~\cite{corrsin1943investigation} and~\cite{wygnanski_fiedler_1969}. Theoretical concepts were published by~\cite{batchelor1948energy}. \cite{george1989self} investigated the properties of various jet and wake flows and showed theoretically that the initial conditions may shape the similarity relations of for instance the round jet in air even in the far field. 

As measurement technology improved, more detailed and accurate measurements were possible, see for example~\cite{panchapakesan_lumley_1993}, who used a hot-wire probe on a moving shuttle and~\cite{hussein_capp_george_1994}, where advanced laser Doppler instruments and “flying hot-wires” were used for the first time. These methods improved the measurements in the outer part of the jet, where high intensity turbulence and back-flow make stationary hot-wire probes inapplicable. More recent measurements pertaining to self-similarity in the free jet can be found in~\cite{burratini_antonia_danaila_2005}, who showed by hot-wire measurements in a large, fully developed jet in air, convincing collapse of scaled statistical parameters including second order.~\cite{ewing_frohnapfel_george_pedersen_westerweel_2007} extended the derivation of similarity variables to two-point statistics, in particular two-point correlations along the jet centerline and showed a similar degree of collapse of scaled two-point correlation functions by measurements with a laser Doppler system upstream and a hot-wire system downstream. Ewing's results have recently been revisited by~\cite{2PointSimRevisited}.

The theory for self-similarity was, as mentioned, initiated by~\cite{zel1937limiting}, and references to the commonly accepted theory for self-similarity in general and for the theory relating to the free axisymmetric jet in particular, to which we shall limit our description in the following, may be found in textbooks, e.g.~\cite{monin2007statistical,tennekes1972first,frisch1995turbulence,pope2000turbulent}. Common for the structure of this theoretical treatment is that the existence of self-similarity is assumed, likely inspired by experiments. The forms of the scaled functions are proposed and inserted into the governing equation, which is taken to be the Navier-Stokes equation in the chosen coordinate system, most often cylindrical coordinates with the z-axis coincident with the jet axis, or some reduced form describing a specific flow, a Reynolds averaged equation or an equation describing only the fluctuating components. The result is then some scaling relations that describe the transformation of the different statistical functions, such as mean velocity profile, second order statistics and higher order statistical functions. With this approach, self-similarity is shown to be possible mathematically, but the existence of self-similarity in real flows must be proven by experiment. 

The existence of self-similarity in the fully developed free axisymmetric jet in the inertial subrange with an initial top-hat profile has been amply proven for the first order statistical quantities such as mean velocity profile and second order static moments. Additionally, the linearly expanding conical shape of the jet, defined e.g. by its half velocity contour, and a virtual origin a few jet diameters downstream from the jet orifice, have been verified by experiments. Careful LDA measurements of second order dynamic moments such as velocity power spectrum, correlation function and second order structure function have nicely supported similarity for these quantities in the inertial subrange, also in parts of the developing region,~\cite{MappingCascadeYaacob}.  

The commonly used turbulent scales such as integral scale, Taylor microscale and Kolmogorov scale also tend to obey a scaling proportional to the distance from the virtual origin, although with some scatter in the value of the origin when extended backwards toward the jet opening. Velocity gradients along the coordinate axes and third order statistical functions that are important for evaluation of the kinetic energy budget are more difficult to measure, and the literature displays considerable scatter in the numerical values. Thus, there are still questions remaining in the literature relating to the number of scaling parameters describing the self-similarity of higher order moments and whether the various statistical functions relate to the same virtual origin,~\cite{thiesset_antonia_djenidi_2014}.

In the present work, we describe a high Reynolds number jet experiment in air using a sophisticated laser Doppler anemometer, developed for velocity measurement in a flow with high turbulence intensity,~\cite{Yaacobetal}. We compare the measured results to the properties expected from the simple model,~\cite{JetSimPart1}, and find that the simple jet model indeed concurs with the obtained measurement results.

\section{\label{sec:TheJetModel}The Jet Model}
\begin{figure}[t]
    \centering
    \includegraphics[width=\linewidth]{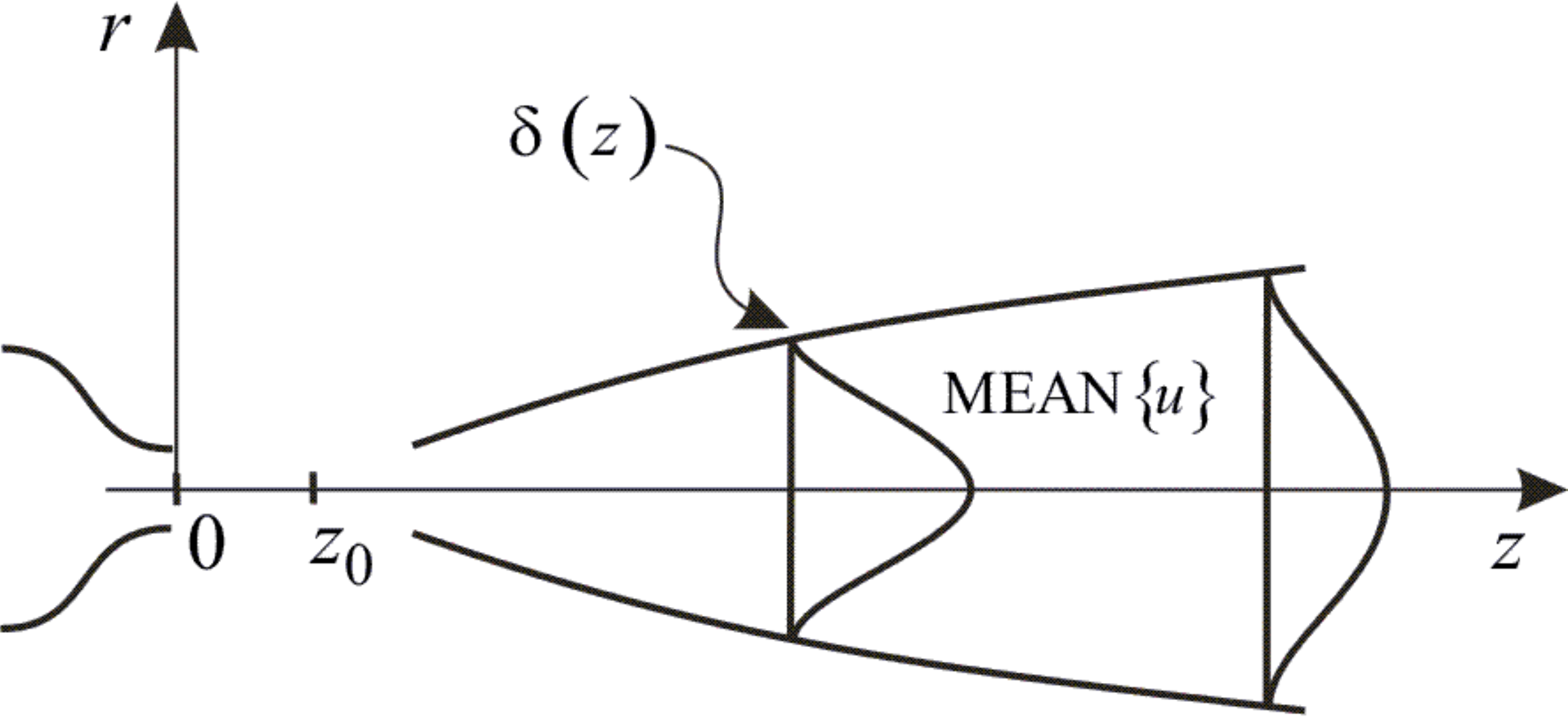}
    \caption{Simple jet model.}
    \label{fig:1}
\end{figure}

In~\cite{JetSimPart1} we describe a simple model of the jet in a cylindrical coordinate system as depicted in Figure~\ref{fig:1}. Initially, nothing is assumed about the shape of the jet contour or the form of the velocity distribution, except that the jet is rotationally symmetric about the jet axis and that the average velocity profiles are integrable. $z$ is the axial distance from the jet exit, the radial coordinate is denoted $r$ and the azimuthal angle is given by $\varphi$. 

The jet width, $\delta$, is in the literature usually defined as the radial distance to a certain fraction of the mean streamwise centerline velocity. A common definition is the jet half width, $\delta_{1/2}(z)$, corresponding to the radial distance from the centerline to half of the mean streamwise centerline velocity. This assumes, of course, a monotonically decreasing mean velocity with increased radial distance from the jet centerline. In the current work, we will apply these definitions, since the velocity profiles are indeed monotonically decreasing with radial distance. However, in the derivation of our simple model,~\cite{JetSimPart1}, we have chosen a different definition to avoid assuming anything about the shape of the velocity profile in order not to implicitly assume self-similarity. Instead, we therein defined the jet width based on a conserved quantity, namely the momentum rate along the $z$-axis. This only assumes integrability of the velocity profiles.

Based on this simple model and the fundamental symmetry relations in classical space and time, the scaling behavior for some static and dynamic first order, second order and third order statistical functions are derived in~\cite{JetSimPart1}. The result of this analysis is listed in Table~\ref{tab:1}, which presents the scaling of the functions along the abscissa and the ordinate relating to plots of the functions in an $r$-$z$-coordinate system. In the following, $u(z,r,\varphi,t)$ designates the time varying instantaneous velocity at the position $(z,r,\varphi)$ at time $t$, and $u'(z,r,\varphi,t)$ is the corresponding fluctuating velocity component. Overbar indicates mean value over a record, whether it be a time record or a spatial record (the choice will be obvious from the context, but in the following, we use exclusively spatial average). Angle brackets indicate ensemble average over a number of records. The combined record and ensemble averaged quantities for the mean, mean square and the variance of the velocity are denoted $\textrm{MEAN}\{ u \}$, $\textrm{MS}\{ u \}$ and $\textrm{VAR}\{ u \}$, respectively. For brevity, we have in several cases suppressed non-affected dependent variables (coordinates) in the formulas in Table~\ref{tab:1}.

\begin{table*}
    \caption{Predicted scaling factors of statistical functions (results from Table 1 in~\cite{JetSimPart1}).}
    \label{tab:1}
    \centering
    \vspace{0.1cm}
    \begin{tabular}{c|c|c|c|c}
    \hline \hline 
    \bf{Statistical Function} & \bf{Definition} & \bf{Dimension} & \bf{Ordinate Scaling Factor} & \bf{Abscissa Scaling Factor} \\ [0.5ex]
    \hline 
    Axial Distance & $z$ & $L$ & $+1$ &  \\
    Jet Width & $\delta (z)$ & $L$ & $+1$ &  \\
    Top Angle & $\theta_0 = \arctan \left ( \frac{\delta (z)}{z} \right ) $ &  &  &  \\
    Mean Velocity & $MEAN\{ u \} = \left \langle \overline{dz / dt} \right \rangle $ & $LT^{-1}$ & $-1$ &  \\
    Mean Square Velocity & $MS\{ u \} = \left \langle \overline{u^2} \right \rangle $ & $L^2T^{-2}$ & $-2$ &  \\
    Velocity Variance & $VAR \{ u \} = \left  \langle \overline{u'^2} \right \rangle $ & $L^2T^{-2}$ & $-2$ &  \\
    Turbulence Intensity & $T = \sfrac{\sqrt{VAR\{u\}}}{MEAN\{u\}}$ &  &  &  \\
    Wave Number & $k$ & $L^{-1}$ & $-1$ & \\
    Power Spectral Density & $F(k) = \left \langle \frac{1}{L_s} \overline{ \left | \hat{u}' (k) \right |^2} \right \rangle $ & $L^3T^{-2}$ & $-1$ & $-1$ \\
    Autocovariance Function & $R(l) = \left \langle \overline{u'(s)u'(s+l)} \right \rangle $ & $L^2T^{-2}$ & $-2$ & $+1$ \\
    Autocorrelation Function & $C(l) = \sfrac{\left \langle \overline{u'(s)u'(s+l)} \right \rangle }{\left \langle \overline{u'(s)^2} \right \rangle} $ &  &  & $+1$ \\
    Integral Length Scale & $I = \int_0^{\infty} C(l) \, dl$ & $L$ & $+1$ & \\
    Taylor Microscale & $\lambda = \sqrt{\sfrac{ \left \langle \overline{u'^2} \right \rangle}{ \left \langle \overline{ \left ( \frac{du'}{dl} \right )^2  } \right \rangle }}$ & $L$ & $+1$ & \\
    Kolmogorov Microscale & $\eta = \left ( \frac{\nu^3}{ \sfrac{\left \langle \overline{|u'|^3} \right \rangle }{  \ell }} \right )^{1/4} $ & $L$ & $+1$ &  \\
    Second Order Structure Function & $S_2(l) = \left \langle \overline{\left ( u(s+l) - u(s) \right )^2 } \right \rangle $ & $L^2T^{-2}$ & $-2$ & $+1$ \\
    Third Order Structure Function & $S_3(l) = \left \langle \overline{\left ( u(s+l) - u(s) \right )^3 } \right \rangle $ & $L^3T^{-3}$ & $-3$ & $+1$ \\
    $S_3$ Slope & $\alpha  =  \left | \frac{d S_3 (l)}{dl} \right | $ & $L^2T^{-3}$ & $-3$ & $+1$ \\
    Mean Dissipation Rate & $\overline{\varepsilon} = \frac{5}{4} \left | \frac{d S_3 (l)}{dl} \right | $ & $L^2T^{-3}$ & $-3$ & $+1$  
    \end{tabular}
\end{table*}

Of special interest is the slope of the third order structure function $S_3(l)$, as it can be used to determine to a good approximation both the average dissipation rate (since the round jet is very close to equilibrium, c.f.~\cite{hussein_capp_george_1994}), $\overline{\varepsilon} = \frac{5}{4} \left | \frac{d S_3 (l)}{dl} \right |$, and the Kolmogorov microscale, $\eta = \left \langle \left ( \frac{\nu^3}{ \overline{\varepsilon}} \right )^{1/4} \right \rangle  $. Here $l$ is the spatial separation in the axial direction between two neighboring measurement points and $\nu$ is the kinematic viscosity. $\ell$ is the pseudo integral length scale. All spatial scales are predicted to grow linearly with $z$, including the Kolmogorov microscale, the spatial Taylor microscale $\lambda = \sqrt{\sfrac{ \left \langle \overline{u'^2} \right \rangle}{ \left \langle \overline{ \left ( \frac{du'}{dl} \right )^2  } \right \rangle }}$, and the integral length scale $I = \int_0^{\infty} C(l) \, dl$, where $C(l)$ is the spatial autocorrelation function. 

In the following section, we describe the LDA measurements of the axial velocity component in a high Reynolds number, axisymmetric jet in air and the calculation of the statistical functions. All calculations provide spatial averages based on the convection of the fluid through the measuring control volume,~\cite{buchhave2017measurement}.

\section{\label{sec:ExpMethod}Experimental Method}
\subsection{Jet Generation Facility}

The jet generator has been designed to produce a top-hat velocity profile at the jet exit using a settling chamber with meshes and baffles transitioning into a nozzle with a smooth fifth order polynomial shape that prevents separation.  The jet had an exit diameter of $D=10\, mm$ and was supplied with air from a pressurized air source. The air pressure was kept constant at $1.0\, bar$, resulting in a jet exit velocity $u_0 = 43\, m/s$ corresponding to a Reynolds number of $32.000$ based on the jet exit diameter and velocity. 

The experiment was conducted inside an enclosure with dimensions $3\,m \times 3\,m  \times 2.5\,m$ (length $\times$ width $\times$ height) to ensure to a good approximation to a free jet flow,~\cite{hussein_capp_george_1994}. A separate pressurized air-line supplied a Laskin-type seeding generator with Glycerine with a pressure of $1.5\, bar$, which was used to seed both the jet and the ambient air in the enclosure. The resulting seeding had a bell-shaped particle size distribution centered around $2\,\mu m$. The seeding generator was operated during at least half an hour prior to a measurement to achieve uniform seed particle distribution throughout the flow.

\subsection{Acquisition}

\begin{figure}[t]
    \centering
    \includegraphics[width=\linewidth]{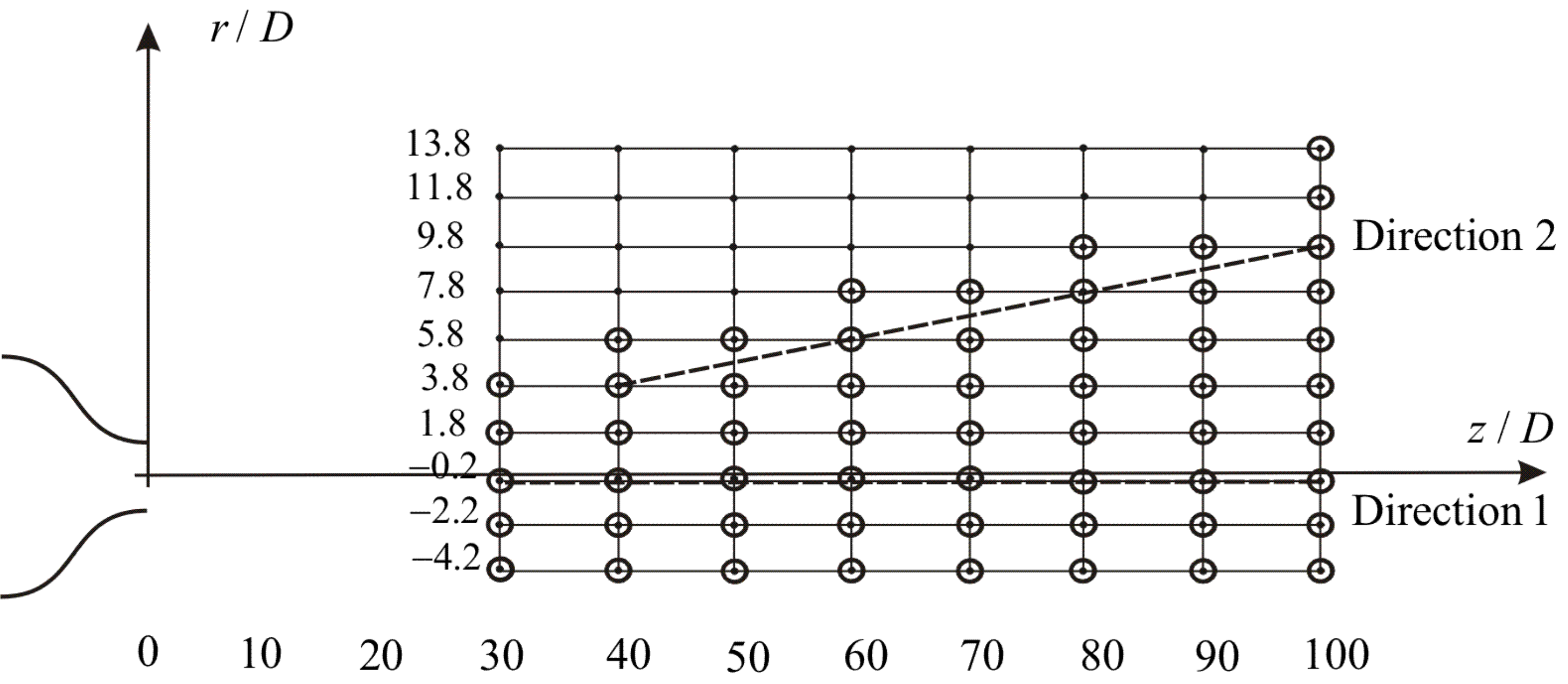}
    \caption{Map of measurement positions; $z$-direction: $30D \rightarrow 100D$ in steps of $10D$. $r$-direction: $-4.2D \rightarrow 13.8D$ in steps of $2D$. The circles indicate points at which data have been acquired.}
    \label{fig:2}
\end{figure}

A map of the measurement positions in a cylindrical coordinate system is shown in Figure~\ref{fig:2} where the circles indicate the acquired measurement points. Direction 1 and Direction 2 refer to measurements following the natural jet development; Direction 1 parallels the jet centerline (with a top angle $\theta_{Dir.\, 1} = 0^{\circ}$) and Direction 2 follows an off-axis development corresponding to the jet half width, $\delta_{1/2} (z)$ (top angle $\theta_{Dir.\, 2} = 5.6^{\circ}$). 

\begin{figure}[t]
    \centering
    \includegraphics[width=\linewidth]{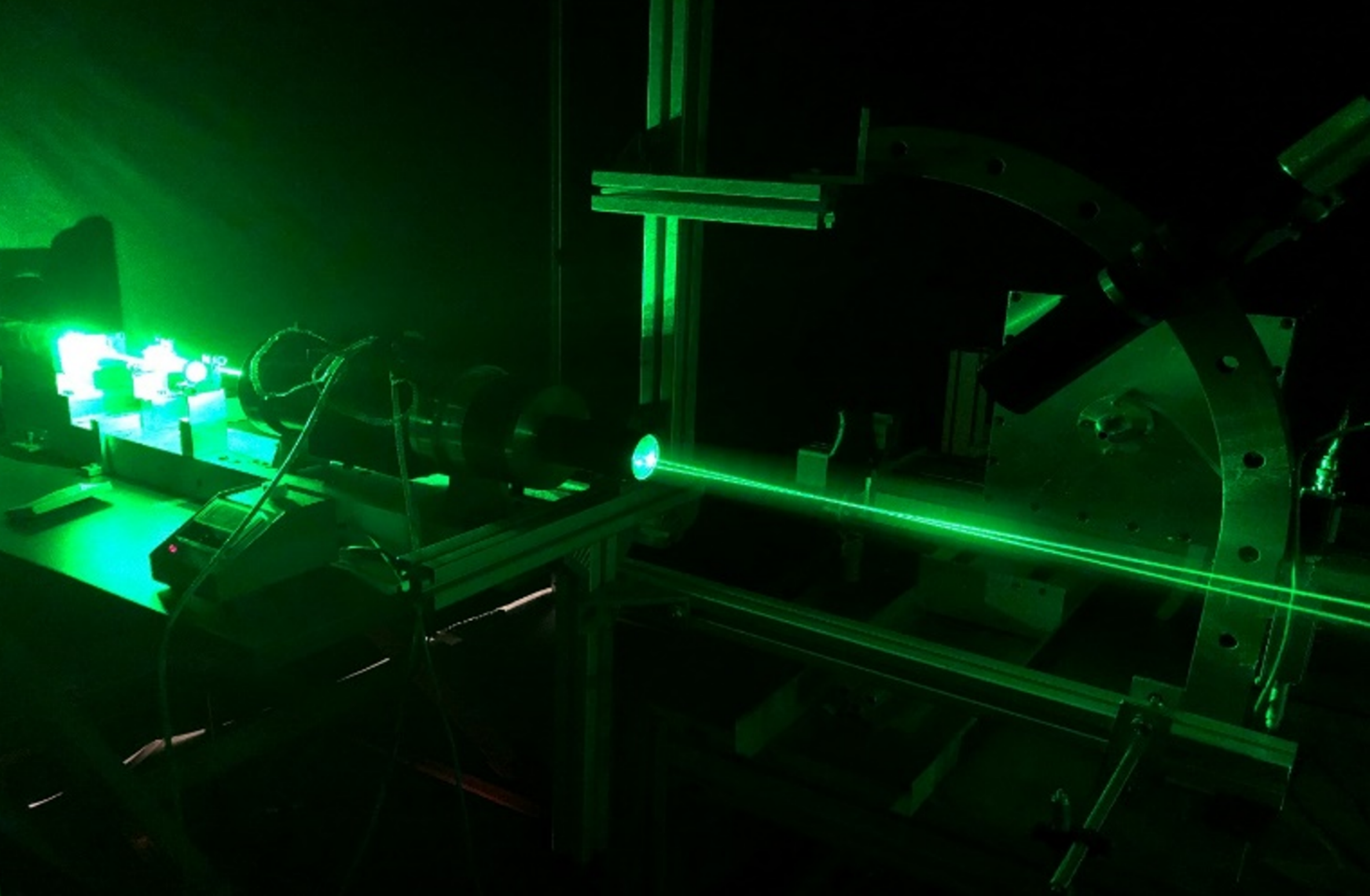}
    \caption{Experimental setup consisting of an in-house designed laser Doppler anemometer (transmitting optics to the left and receiving optics to the upper right) and an axisymmetric jet generation facility (right).}
    \label{fig:3}
\end{figure}

The axial velocity component was measured with an in-house designed laser Doppler anemometer (\cite{Yaacobetal}), see Figure~\ref{fig:3}. The transmitting optics was mounted next to the jet aligned to measure the axial velocity component. The detecting optics was mounted in a forward scattering off-axis direction with a scattering angle of $45^{\circ}$. For maximum stability, the LDA was kept at a fixed position while the jet was traversed. 

The detector was a Hamamatsu photomultiplier (type H10425) with a load resistor of $270\,\Omega$ allowing an electronic bandwidth of $10\, MHz$. The photomultiplier was followed by a low noise preamplifier (FEMTO HVA-200M-40-F) and the signal was digitized with a 12-bit digital oscilloscope (Picoscope 5000). The digitized signal was transferred to computer memory or to an external hard disk. 

At each measurement position, 100 – 400 records (number increasing towards the far field, each 1 – 4 seconds long (again increasing towards the far field), were measured. The sample rate of the digital oscilloscope was adjusted to between 2-4 times the Nyquist rate for the frequency shifted Doppler signal. The optical frequency shift is selected for each measurement point to shift the received frequency modulation to a range, which is optimal for applying post detection band-pass filtering in order to obtain optimal signal-to-noise ratio and minimum electronic bias of the received Doppler bursts. The frequency shift was generated by a dual-Bragg cell module (two IntraAction AOM-402AF1 with dual-RF driver DFE-404A4) and the difference frequency shift could be freely adjusted between $0$ and $4\, MHz$.

Several features distinguish this LDA and make it ideal for measurements in a highly turbulent flow, see also~\cite{Yaacobetal}. The off-axis configuration allows a very small, nearly spherical measuring volume with a diameter of $140\, \mu m$. The adjustable frequency shift makes it possible to choose the best frequency range for each measurement position for accurate, bias-free measurements in a highly turbulent flow, including backflow. The low noise amplification with matching bandwidth and 12-bit digitization helps to achieve a signal with a high signal-to-noise ratio.

\subsection{Signal and Data Processing}\label{subsec:signdatproc}

The signal processing is software based, see~\cite{Yaacobetal}. Upon processing, each measured (burst) signal time record results in a data file containing measured values for particle arrival time,  residence time, and velocity. In this section, we introduce the formulas used in the numerical calculations of statistical functions. We use the following definition for the constants and variables:\\

\noindent $n$ : sample index \newline 
\noindent $m$ : record index \newline 
\noindent $t_n$ : sampling time \newline 
\noindent $s_n$ : sampling length (mapped spatial record) \newline 
\noindent $\omega_n$ : temporal frequency \newline 
\noindent $k_n$ : spatial wavenumber \newline 
\noindent $u_n$ : instantaneous velocity sample \newline 
\noindent $\Delta s_n$ : spatial residence length \newline 
\noindent $\Delta t_n$ : residence time \newline 
\noindent $N$ : number of samples in a record \newline 
\noindent $M$ : number of acquired records \newline 
\noindent $\nu$ :  kinematic viscosity \newline 

The measured residence time in connection with the known measuring volume diameter allows the conversion of the measured temporal record to a spatial record, the so-called convection record,~\cite{buchhave2017measurement}. With this method, the sweeping effect due to the fluctuating convection velocity is removed, and the record displays the spatial velocity structures as they are convected past the measuring volume. As explained in~\cite{buchhave2017measurement}, a time record from a stationary flow phenomenon is by this method converted to a homogeneous spatial record, allowing single point calculations of spatial statistical functions. In the case of an LDA measurement, this only requires that the seed particles are uniformly (albeit randomly) distributed in space. 

For LDA systems operating in burst-mode, time averaging must be correctly performed using residence time corrected algorithms (so-called residence time weighting), see e.g.~\cite{AnnurevBGL,EstimationBmodeLDA}. However, spatial averaging does not require sample rate bias correction and must be carried out using arithmetic averaging. 

In the following, unless time average is specifically addressed, all averaging operations and calculations will apply arithmetic averaging to spatial records. We obtain the final averaged statistics by subsequent ensemble averaging over all the measured records.

Overbar denotes average over $N$ samples in a record and angle brackets indicate ensemble average over all $M$ records.\\

\noindent \textbf{Static Moments:}\\

\noindent \textrm{Mean Velocity: }
\begin{equation*} 
\begin{split}
\overline{u} & = \frac{1}{N} \sum_{n=0}^{N-1} u_n,\\  
\textrm{MEAN} \{u\} & = \left \langle \overline{u} \right \rangle 
\end{split}
\end{equation*}

\noindent \textrm{Fluctuating Velocity: }
\begin{equation*} 
\begin{split}
u' & = u -\overline{u}
\end{split}
\end{equation*}

\noindent \textrm{Mean Square Velocity: }
\begin{equation*} 
\begin{split}
\overline{u^2} & = \frac{1}{N} \sum_{n=0}^{N-1} u_n^2 \\ 
\textrm{MS} \{u\} & = \left \langle \overline{u^2} \right \rangle 
\end{split}
\end{equation*}

\noindent \textrm{Velocity Variance:}
\begin{equation*} 
\begin{split}
\overline{u'^2} & = \frac{1}{N} \sum_{n=0}^{N-1} u_n'^2 \\
\textrm{VAR} \{u\} & = \left \langle \overline{u'^2} \right \rangle 
\end{split}
\end{equation*}\\

\noindent \textbf{Dynamic Moments:}\\

\noindent The data calculations are, as far as possible, performed by array processing. This allows Fourier transforms to be computed from the unaliased, randomly sampled velocity data by means of the discrete Fourier transform (DFT) using, in the current case, an exponentially increasing frequency scale. This results in equally spaced points when the power spectrum is displayed in a logarithmic graph. Hence, far fewer spectral values need to be computed than when a fast Fourier transform (FFT) is used, in which case the Nyquist criterion must be adhered to in order to avoid aliasing.\\

\noindent \textrm{Spatial Fourier Transform: }
\begin{equation*} 
\begin{split}
\hat{u}'(k) & = \sum_{n=0}^{N-1} e^{-iks_n}u_n' \Delta s_n
\end{split}
\end{equation*}

\noindent \textrm{Power Spectral Density (PSD): }
\begin{equation*} 
\begin{split}
F(k)_m & = \frac{1}{L_s} \hat{u}'(k) \hat{u}'(k)^{\ast}\\
F(k) & = \left \langle F(k)_m \right \rangle
\end{split}
\end{equation*}
where $L_s$ is the spatial record length.

Autocovariance functions (ACF) and autocorrelation functions (normalized ACF) are computed by forming velocity products with lags created from the available randomly sampled data. Note that we use the term lag for both temporal distances, $\tau$, and spatial separations along the convection record between data points, $l$. We also use the term `frequency' for both temporal frequency and spatial frequency. Due to the inherent random sampling of the LDA, the velocity products are collected in equidistant slots of width $\Delta$ and averaged by normalizing each slot by the number of products obtained, $N_i$ (the so-called slotted autocovariance function, SACF). 

The measured velocities contain a random noise that is apparent in the statistics when a velocity value is correlated with itself (and averages out when velocity samples are correlated with other samples than themselves). Thus, at zero lag, both the measured velocities and the random noise are fully correlated, whilst for non-zero lags the random noise correlations vanish for converged statistics. The zero lag value (or self-products) thus contain high frequency measurement noise not connected to the turbulent velocity, which manifests itself as a `spike' at zero lag. To avoid this noise contribution from the self-products of the velocities, we have formed the zero lag value by second order extrapolation of the first few slots back to zero lag with a horizontal tangent at $l=0$.\\

\noindent \textrm{Covariance function: }
\begin{equation*} 
\begin{split}
R(l_i) & = \frac{1}{N_i} \sum_{n \neq n'} \left ( u'(s_n) \cdot u'(s_{n'}) \right ), \,\, l_i - \frac{\Delta}{2} < s_n - s_{n'} < l_i + \frac{\Delta}{2}
\end{split}
\end{equation*}

\noindent \textrm{Correlation function: }
\begin{equation*} 
\begin{split}
C(l_i) & = R(l_i) / R(0)
\end{split}
\end{equation*}
where $N_i$ is the number of product contributions (realizations) in slot $i$ and the double sum covers all samples (except self-products $n=n'$) for the running parameters $n$ and $n'$. 

Spatial structure functions are likewise formed from velocities displaced in the spatial convection record direction with the actual measured lag values. The velocity differences for each slot are then squared and averaged in case of the second order structure function or elevated to the third power and averaged in case of the third order structure function. Again, we apply the slotting technique and normalize with the number of realizations obtained into each slot, $N_i$. We assume the turbulence is in equilibrium (which is a reasonable approximation for the axisymmetric jet, see e.g.~\cite{hussein_capp_george_1994}), so we can apply Kolmogorov's $4/5-$law and compute the average dissipation rate from the initial slope of the third order structure function. We can then estimate the Kolmogorov microscale from the dissipation and the kinematic viscosity of air.\\

\noindent \textrm{Second order structure function: }
\begin{equation*} 
\begin{split}
S_2(l_i)_m & = \frac{1}{N_i} \sum_{n \neq n'} \left ( u(s_n) - u(s_{n'}) \right )^2, \,\, l_i - \frac{\Delta}{2} < s_n - s_{n'} < l_i + \frac{\Delta}{2}\\
S_2(l_i) & = \left \langle S_2(l_i)_m \right \rangle
\end{split}
\end{equation*}

\noindent \textrm{Third order structure function: }
\begin{equation*} 
\begin{split}
S_3(l_i)_m & = \frac{1}{N_i} \sum_{n \neq n'} \left ( u(s_n) - u(s_{n'}) \right )^3, \,\, l_i - \frac{\Delta}{2} < s_n - s_{n'} < l_i + \frac{\Delta}{2}\\
S_3(l_i) & = \left \langle S_3(l_i)_m \right \rangle
\end{split}
\end{equation*}

\noindent \textrm{Third order structure function slope: }
\begin{equation*} 
\begin{split}
\alpha & = \left | \frac{dS_3 (l)}{dl} \right |_{l=0}\mathrm{,} 
\end{split}
\end{equation*}
where the derivative is performed digitally.\\

\noindent \textbf{Spatial Scales:}\\

\noindent \textrm{Average dissipation rate: }
\begin{equation*} 
\begin{split}
\overline{\varepsilon} & = \frac{5}{4} \left | \alpha \right | =   \frac{5}{4} \left | \frac{dS_3 (l)}{dl} \right |_{l=0} 
\end{split}
\end{equation*}

The spatial Kolmogorov microscale is computed from the average dissipation rate by $\eta = \left ( \frac{\nu^3}{\overline{\varepsilon}} \right )^{1/4}$. Using the slope from the third order structure function we obtain the Kolmogorov microscale. \\

\noindent \textrm{Kolmogorov microscale: }
\begin{equation*} 
\begin{split}
\eta & = \left ( \frac{4 \nu^3}{5 \left | \alpha \right |} \right )^{1/4}
\end{split}
\end{equation*}

The spatial integral scale is computed from the integral of the spatial correlation function, $I \equiv \int_0^{\infty}C(l) \, dl$. As the measured records are very long compared to the integral scale, we have, in order to reduce noise, limited the upper limit of the integral to lags with a clear correlation signature, corresponding to two integral length scales, $l_{2I}$. This corresponds in the discrete case to an upper limit sample number in the sum of $n_{2I}$. \\

\noindent \textrm{Integral length scale: }
\begin{equation*} 
\begin{split}
I & = \sum_{n=0}^{n_{2I}} C(l_n)
\end{split}
\end{equation*}

The square of the Taylor microscale is defined as the ratio of the mean square fluctuating velocity divided by the mean square of the derivative of the fluctuating velocity, 
$$\lambda^2 \equiv \frac{\left \langle \overline{u'(s)^2} \right \rangle}{\left \langle  \overline{\left ( du'(s)/ds \right )^2} \right \rangle}.$$ 
However, having available the Fourier components of the velocity field, we have chosen to compute the Taylor scale in $k$-space:\\

\noindent \textrm{Taylor microscale: }
\begin{equation*} 
\begin{split}
\lambda^2 & \equiv  \frac{\left \langle \overline{\hat{u}'(k)\hat{u}'(k)^{\ast}} \right \rangle}{\left \langle  \overline{k^2 \hat{u}'(k)\hat{u}'(k)^{\ast} } \right \rangle}
\end{split}
\end{equation*}

As the spatial Taylor microscale is very dependent on the wavenumber range, we have, to obtain a well-defined value for the Taylor scale, based the estimates on the data in the wavenumber range with a $-5/3$ slope, the inertial sub-range. This is simply done by including only Fourier components in this frequency range in the calculation. \\

\noindent \textrm{Taylor microscale: }
\begin{equation*} 
\begin{split}
\lambda_m & =  \frac{\sqrt{\sum_{n=n_{5/3}}\hat{u}'(k_n)\hat{u}'(k_n)^{\ast}}}{\sqrt{\sum_{n=n_{5/3}}k_n^2 \, \hat{u}'(k_n)\hat{u}'(k_n)^{\ast}}}, \quad \quad \lambda = \left \langle \lambda_m \right \rangle
\end{split}
\end{equation*} 

\section{\label{sec:Results}Results}

\subsection{Static moments}

\begin{figure}[t]
    \begin{minipage}{\linewidth}
    \centering
    \includegraphics[width=\linewidth]{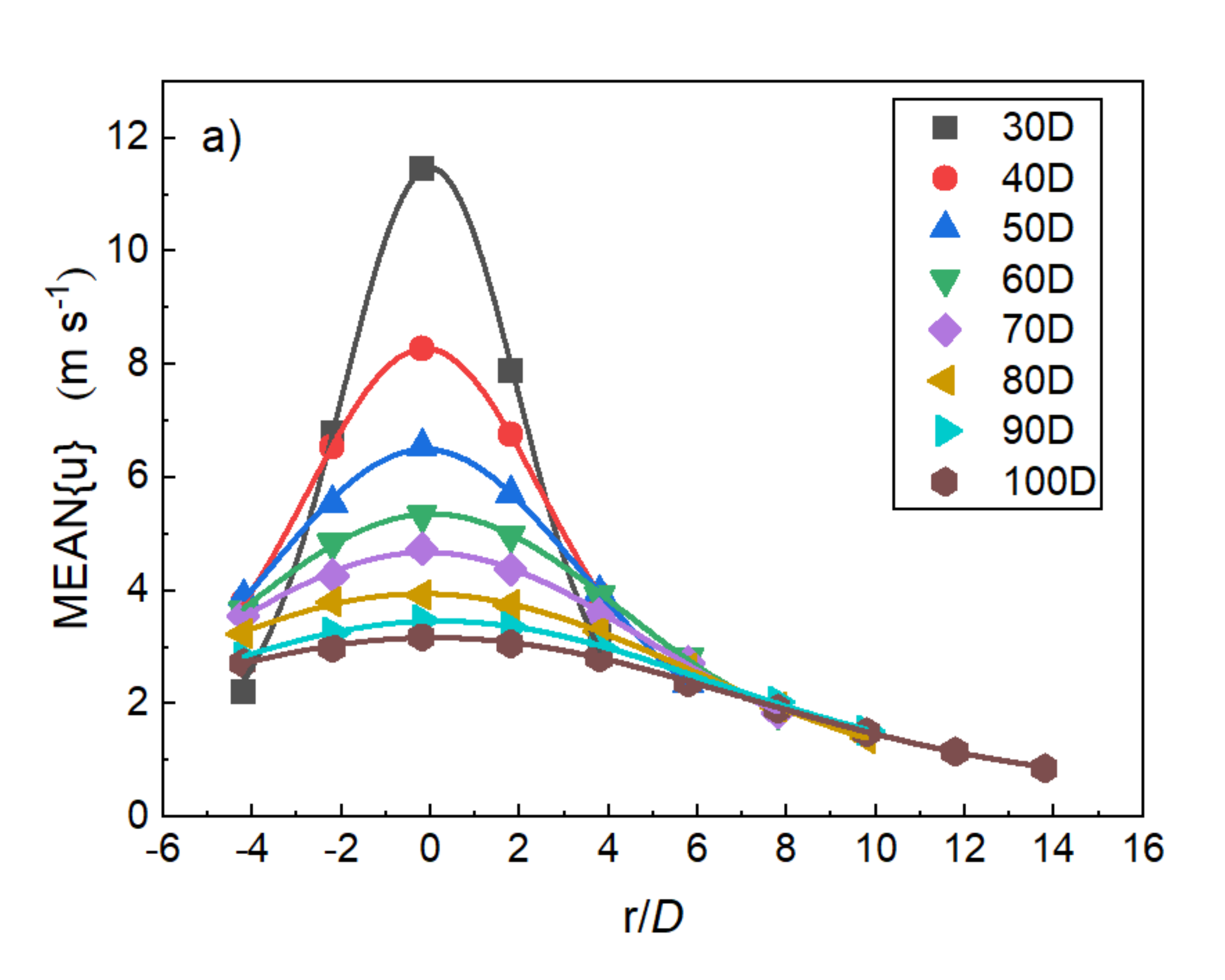}
    \end{minipage}
        \begin{minipage}{\linewidth}
    \centering
    \includegraphics[width=\linewidth]{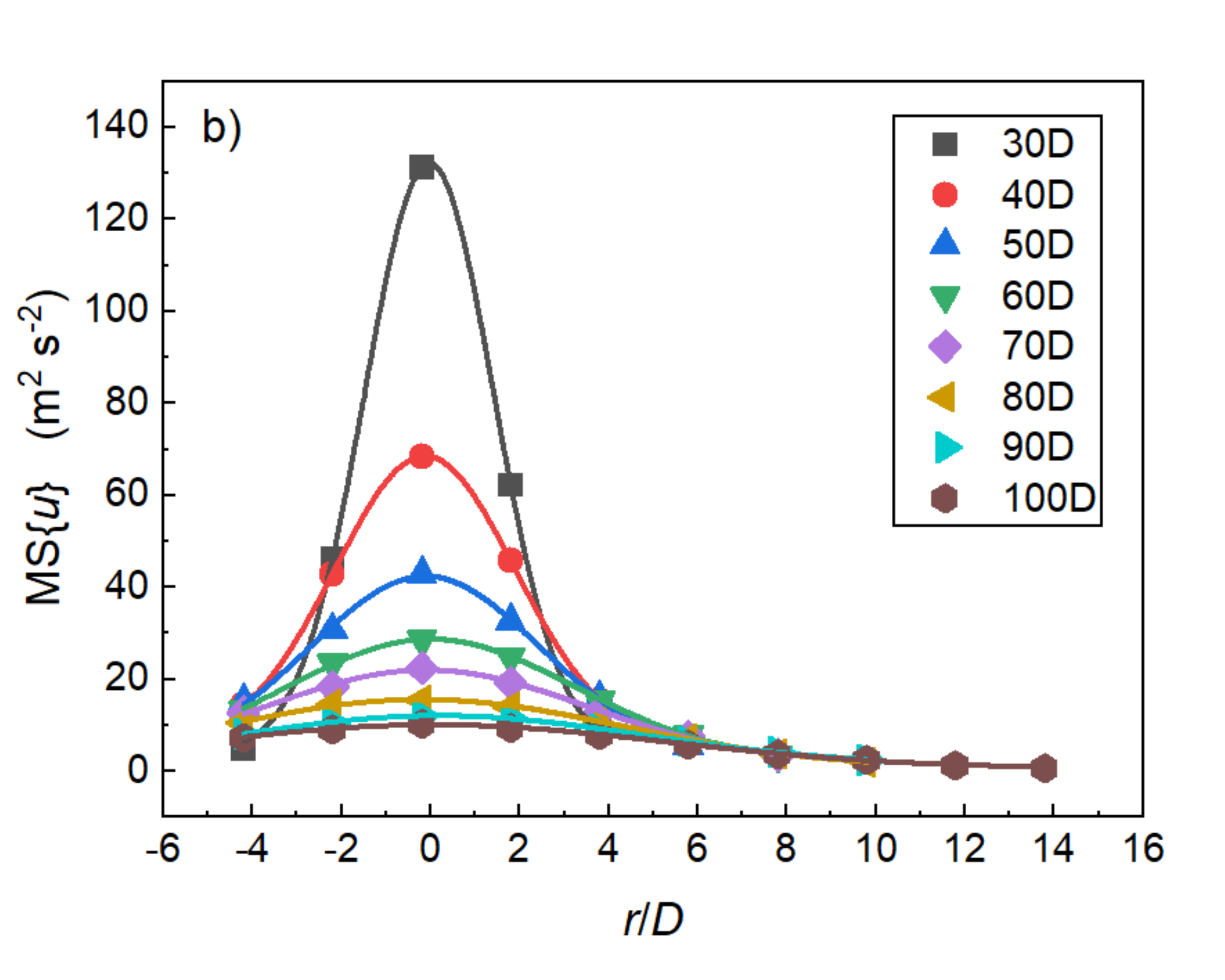}
    \end{minipage}
    \caption{Transverse profiles of a) mean streamwise velocity; b) mean squared streamwise velocity along the downstream direction $z/D$. Each curve is fitted with a Gaussian distribution.}
    \label{fig:4}
\end{figure}

Time records of the axial velocity component were measured in selected positions (as indicated by circles in Figure~\ref{fig:2}) in the jet within a matrix covering the axial locations $z/D = 30 : 10 : 100$ and transverse locations $r/D = -4.2 : 2 : 13.8$. Figure~\ref{fig:4}$a$ shows the measured radial profiles of the mean axial velocity component. The measured values are fitted with Gaussian distributions with standard deviation $\sigma$. Figure~\ref{fig:4}$b$ shows the corresponding mean square profiles of the axial velocity, again fitted with Gaussian distributions. 

\begin{figure}[t]
    \begin{minipage}{\linewidth}
    \centering
    \includegraphics[width=\linewidth]{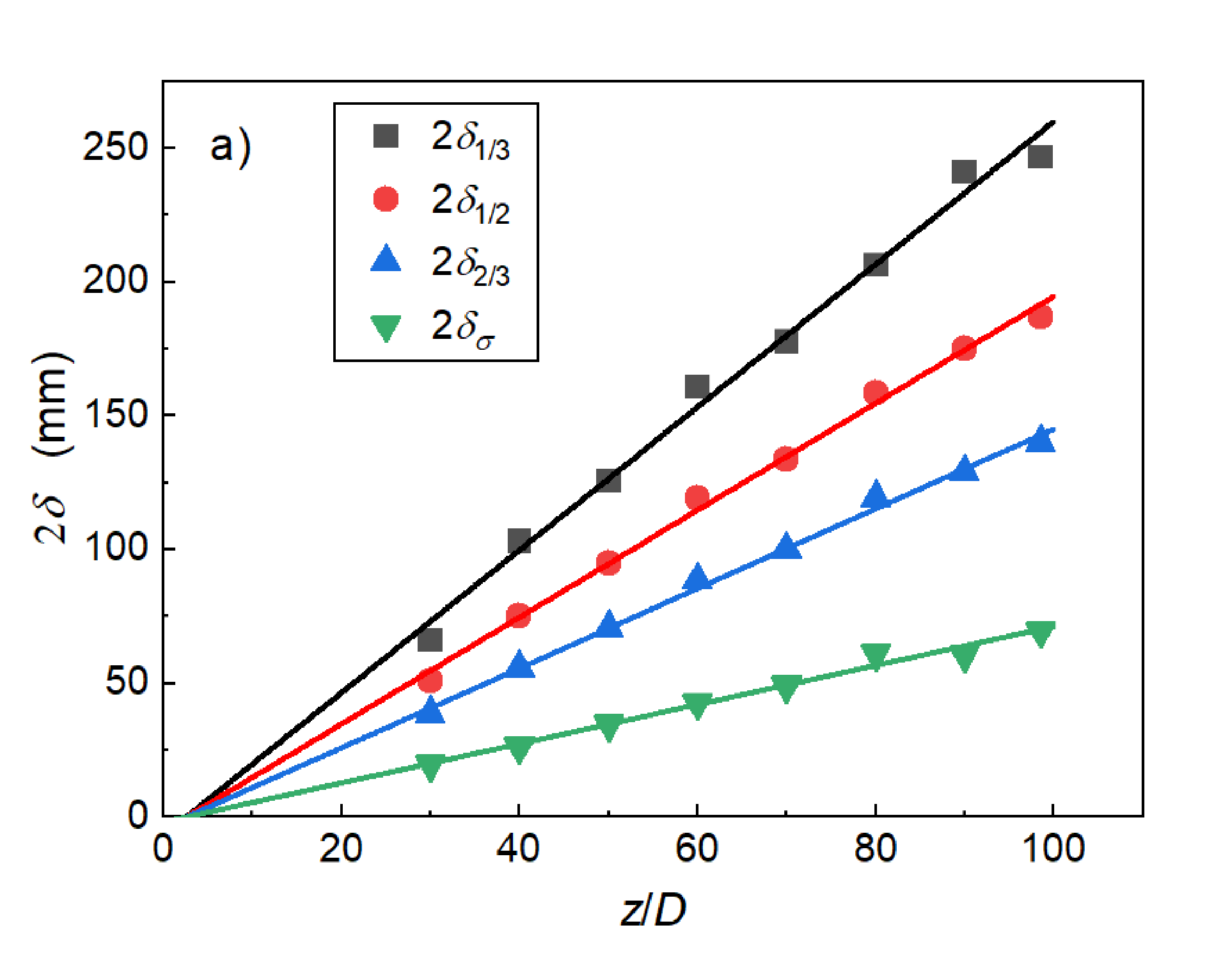}
    \end{minipage}
        \begin{minipage}{\linewidth}
    \centering
    \includegraphics[width=\linewidth]{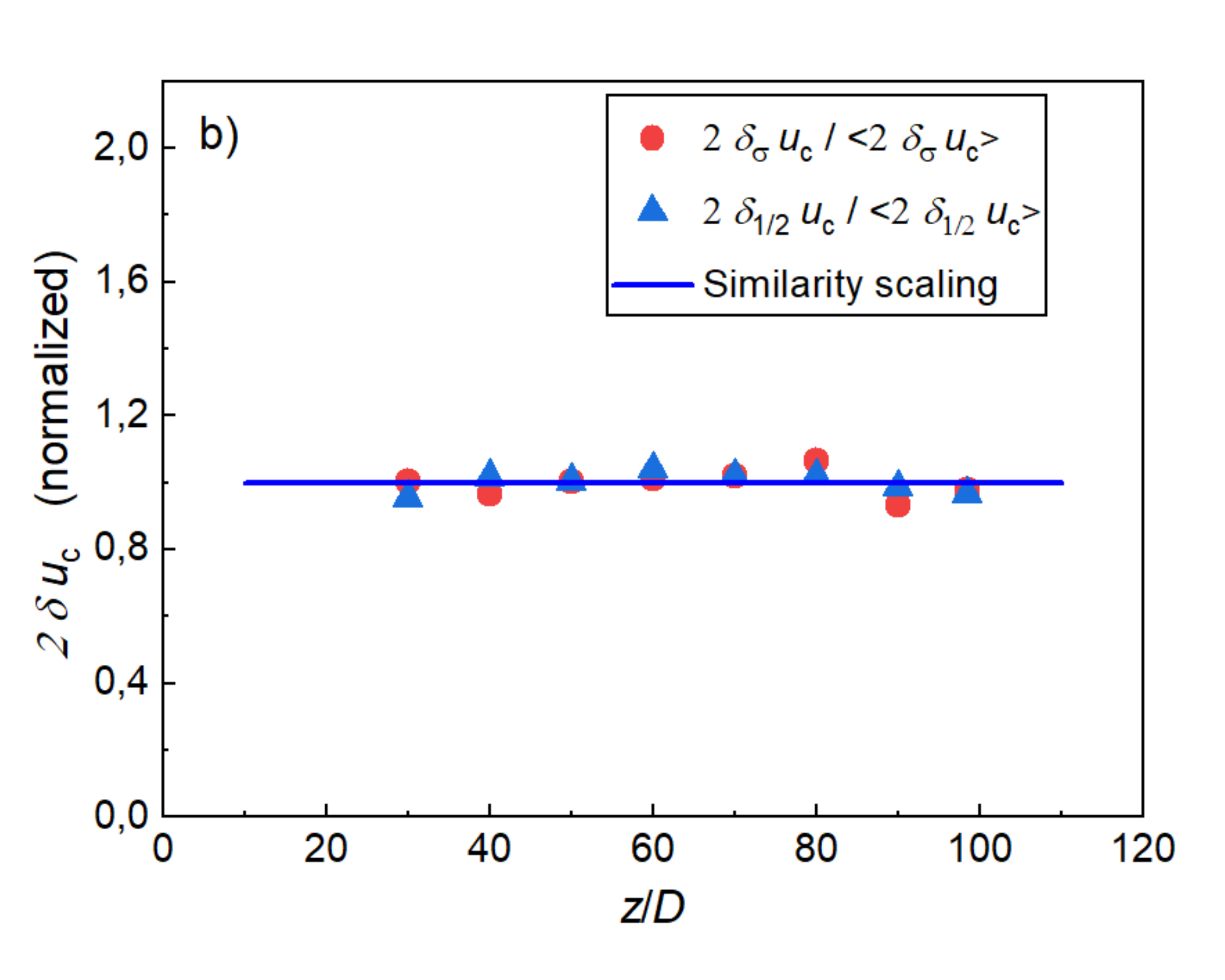}
    \end{minipage}
    \caption{Linear expansion of the jet: a) Full width at different heights (1/3, 1/2, 2/3 of the centerline velocity, $u_c$, and full width of the jet at $\sigma$, $2\delta_{\sigma}$, of the Gaussian fit), along the downstream distance. Lines indicate self-similar first order polynomia. b) Normalized products of $2\delta_{1/2}$ and $2\delta_{\sigma}$ with $u_c$, respectively. }
    \label{fig:5}
\end{figure}

Figure~\ref{fig:5}$a$ shows the jet full width at different fractions of the mean axial centerline velocity; $2\delta_{1/3}$, $2\delta_{1/2}$, $2\delta_{2/3}$ and $2\delta_{\sigma}$. The measurements for each definition of the jet width is fitted to a first order polynomial intersecting the $z$-axis at the virtual origin, $z_0$, as expected from the predicted similarity scaling in~\cite{JetSimPart1} (in the following referred to as ``Similarity scaling'' in the figure legends). A value of $z_0/D=5$ is seen to represent the virtual origin for the different jet contours quite well.

Figure~\ref{fig:5}$b$ shows the products of the mean centerline velocity, $u_c$, and the full width at half height (FWHH) of the jet, $2\delta_{1/2}$, as well as the product of $u_c$ and the full width of the jet at the sigma value of the Gaussian fit, $2\delta_{\sigma}$, normalized by the average product along the jet. This product thus represents the product of the large characteristic spatial scale and the large characteristic velocity scale and thereby the constant momentum flux along the axis and thereby indicates a constant large scale Reynolds number and thus self-similarity of the jet.

\begin{figure}[t]
    \begin{minipage}{\linewidth}
    \centering
    \includegraphics[width=\linewidth]{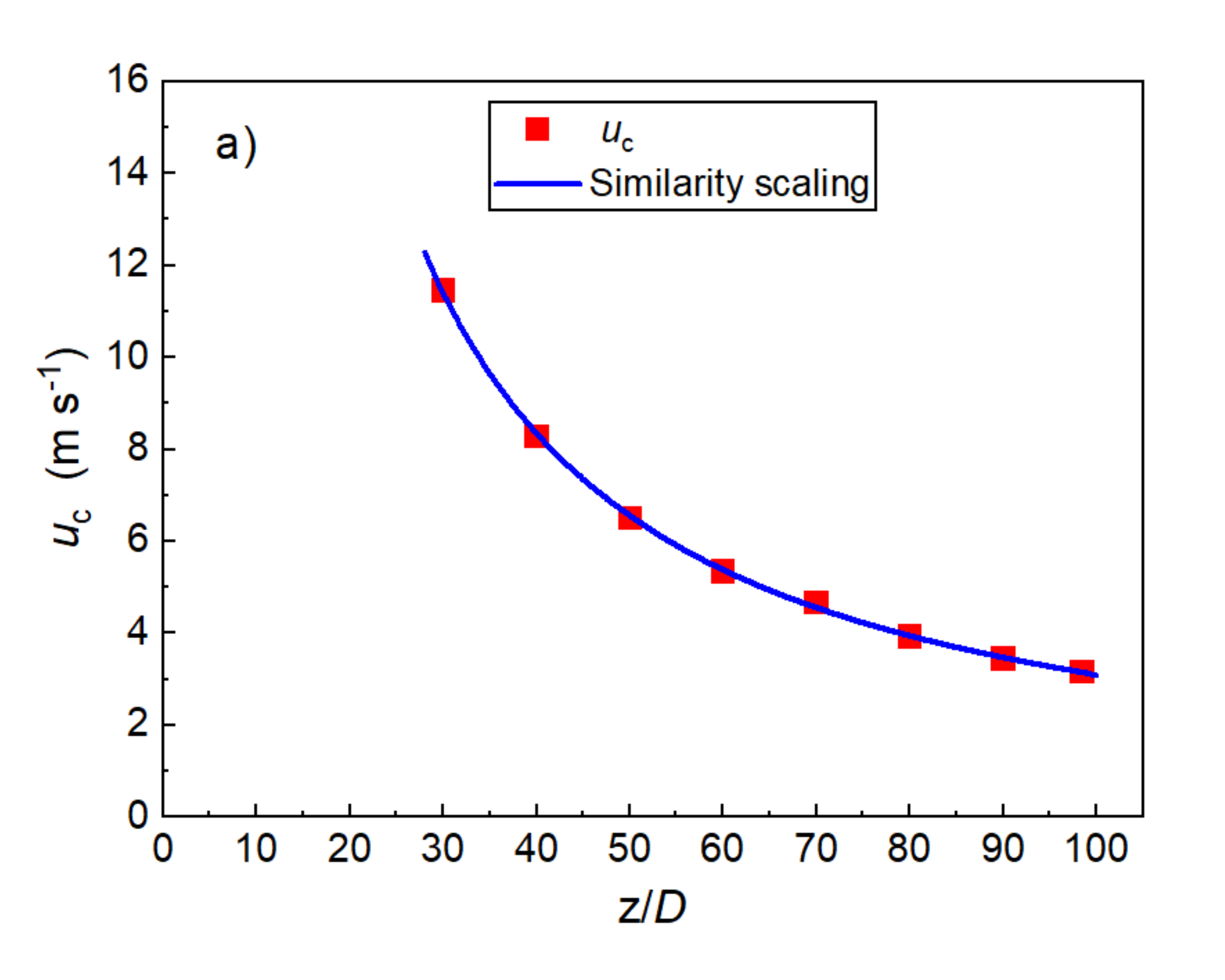}
    \end{minipage}
    \begin{minipage}{\linewidth}
    \centering
    \includegraphics[width=\linewidth]{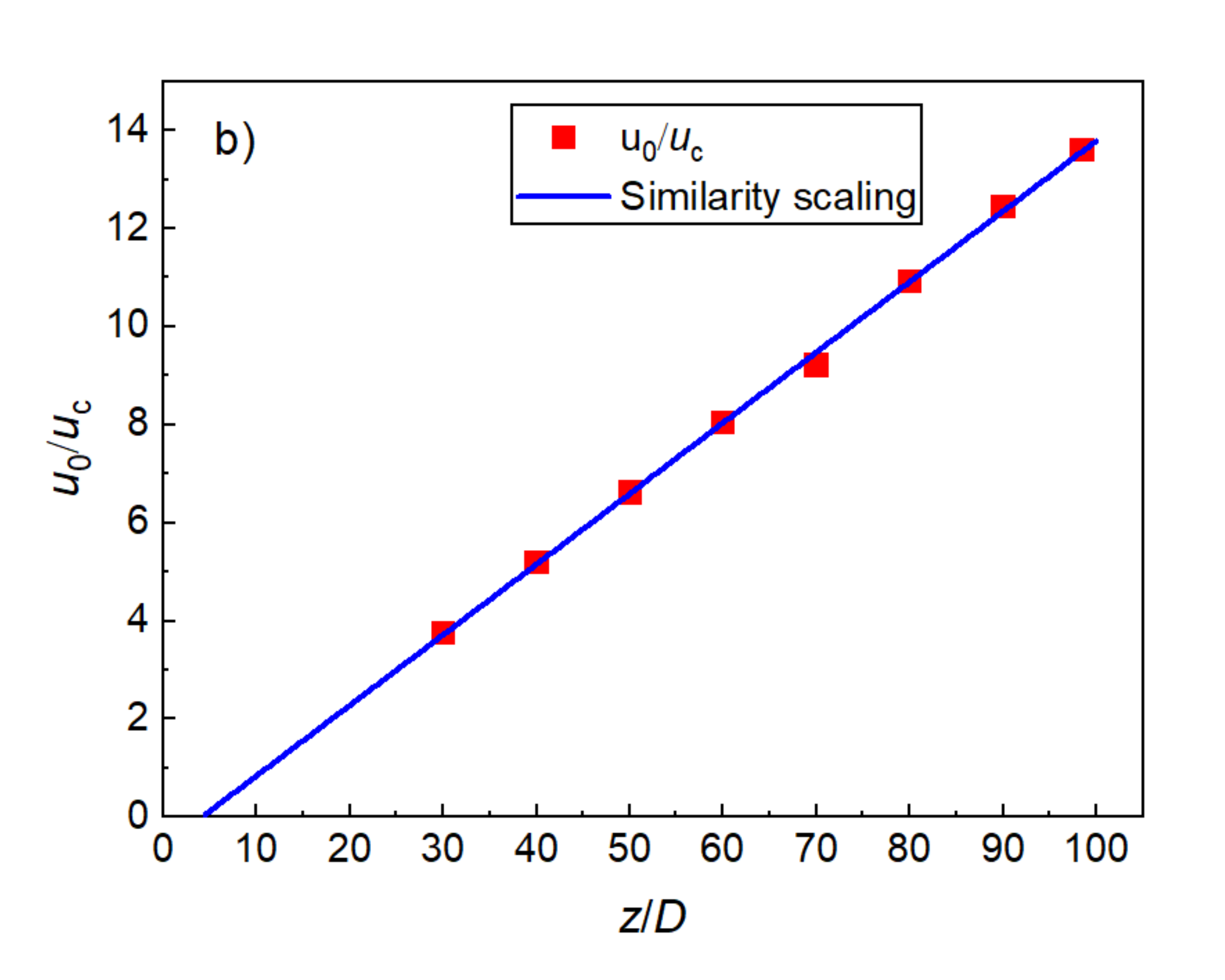}
    \end{minipage}
    \caption{Downstream development of a) the mean axial centerline velocity, $u_c$, and b) the inverse of the mean axial centerline velocity normalized by the jet exit velocity, $u_0=43\,m/s$. The blue lines express the expected respective similarity scaling ($\propto x^{-1}$ and $\propto x$), from~\cite{JetSimPart1}.}
    \label{fig:6}
\end{figure}

Figure~\ref{fig:6}$a$ shows the mean axial centerline velocity, $u_c$, as a function of the downstream distance, $z$. The line is an inverse function of the axial distance from the virtual origin, $z-z_0$ as expected from self-similarity. Figure~\ref{fig:6}$b$ shows the inverse mean axial centerline velocity normalized by the jet exit velocity, $u_0$, as a function of $z$. The virtual origin $z_0$ again appears to lie around $5D$. The lines in Figures~\ref{fig:6}$a$ and $b$ are  not fitted to the measurement data, but instead constitute first order functions of the distance from the virtual origin, as expected from the simple model in~\cite{JetSimPart1}.

\begin{figure}[t]
    \begin{minipage}{\linewidth}
    \centering
    \includegraphics[width=\linewidth]{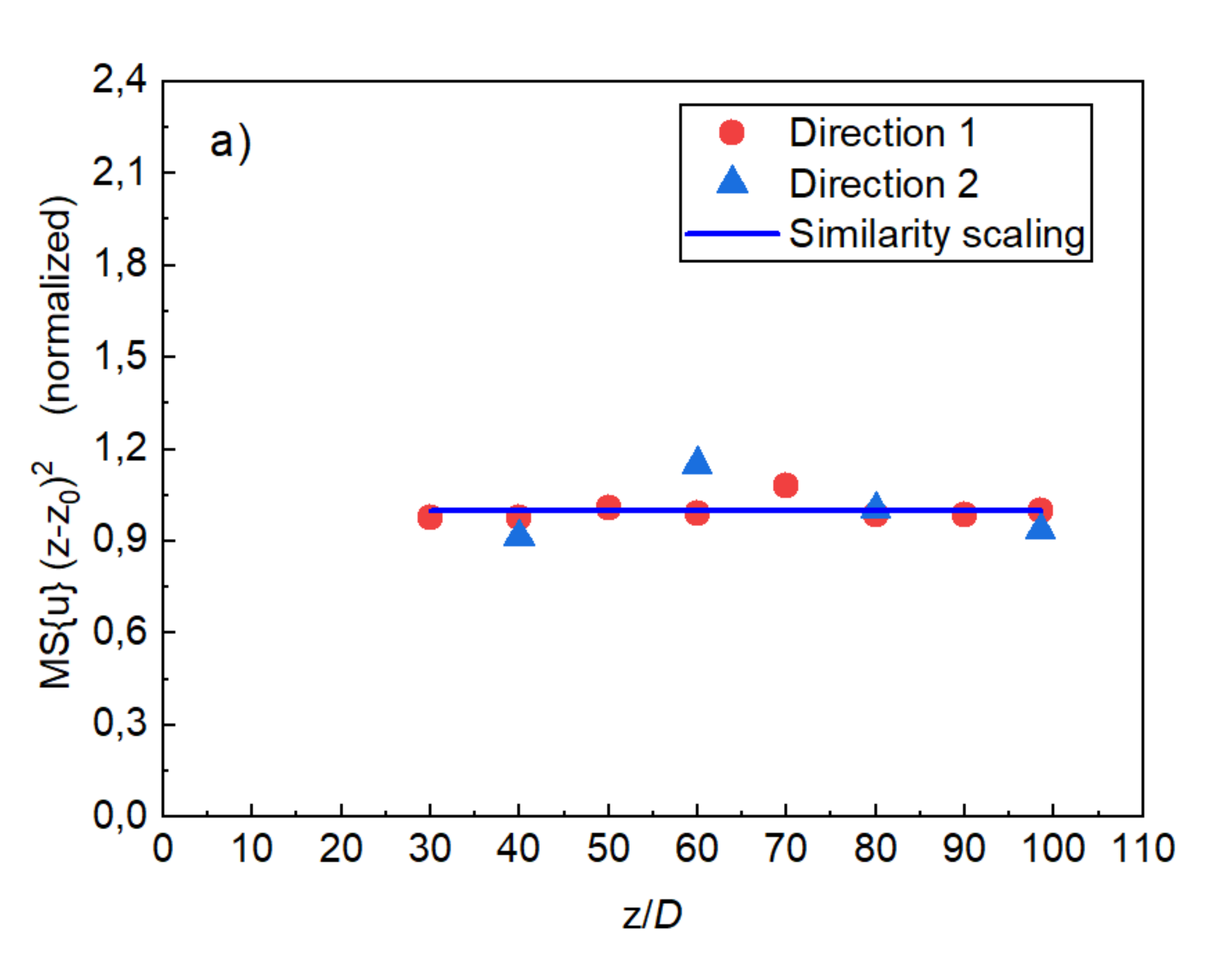}
    \end{minipage}
    \begin{minipage}{\linewidth}
    \centering
    \includegraphics[width=\linewidth]{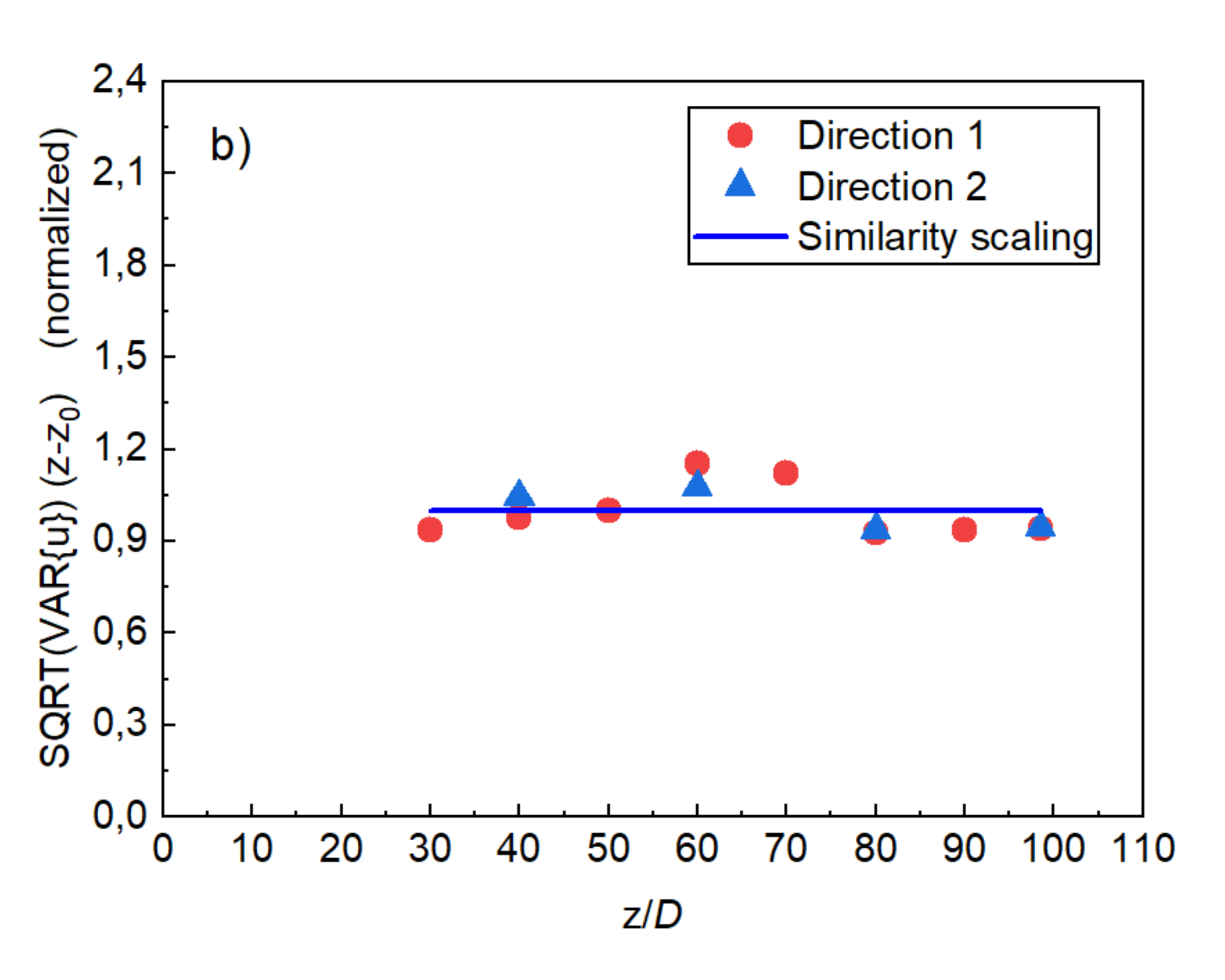}
    \end{minipage}
    \caption{a) Normalized product of the mean square velocity and $(z-z_0)^2$ and b) normalized product of RMS velocity and $(z-z_0)$ in the fully developed region ($30\leq z/D \leq 100$) of the jet along Direction 1 and Direction 2, respectively.}
    \label{fig:7}
\end{figure}

Figure~\ref{fig:7}$a$ and $b$ show along Directions 1 and 2, respectively, the mean square (MS) and the root-mean-square (RMS) of the axial velocity component. These quantities have been scaled by multiplication by the distance from the virtual origin $(z-z_0)$ (for the RMS) and, by the square of this distance, $(z-z_0)^2$ (for the MS), in accordance with the expected similarity scaling in~\cite{JetSimPart1}. The constant values of the products thus indicate self-similarity along both Directions 1 and 2. These products are, in turn, normalized by their average value along the axis to scale the values so they are distributed around a value of unity. 

\begin{figure}[t]
    \centering
    \includegraphics[width=\linewidth]{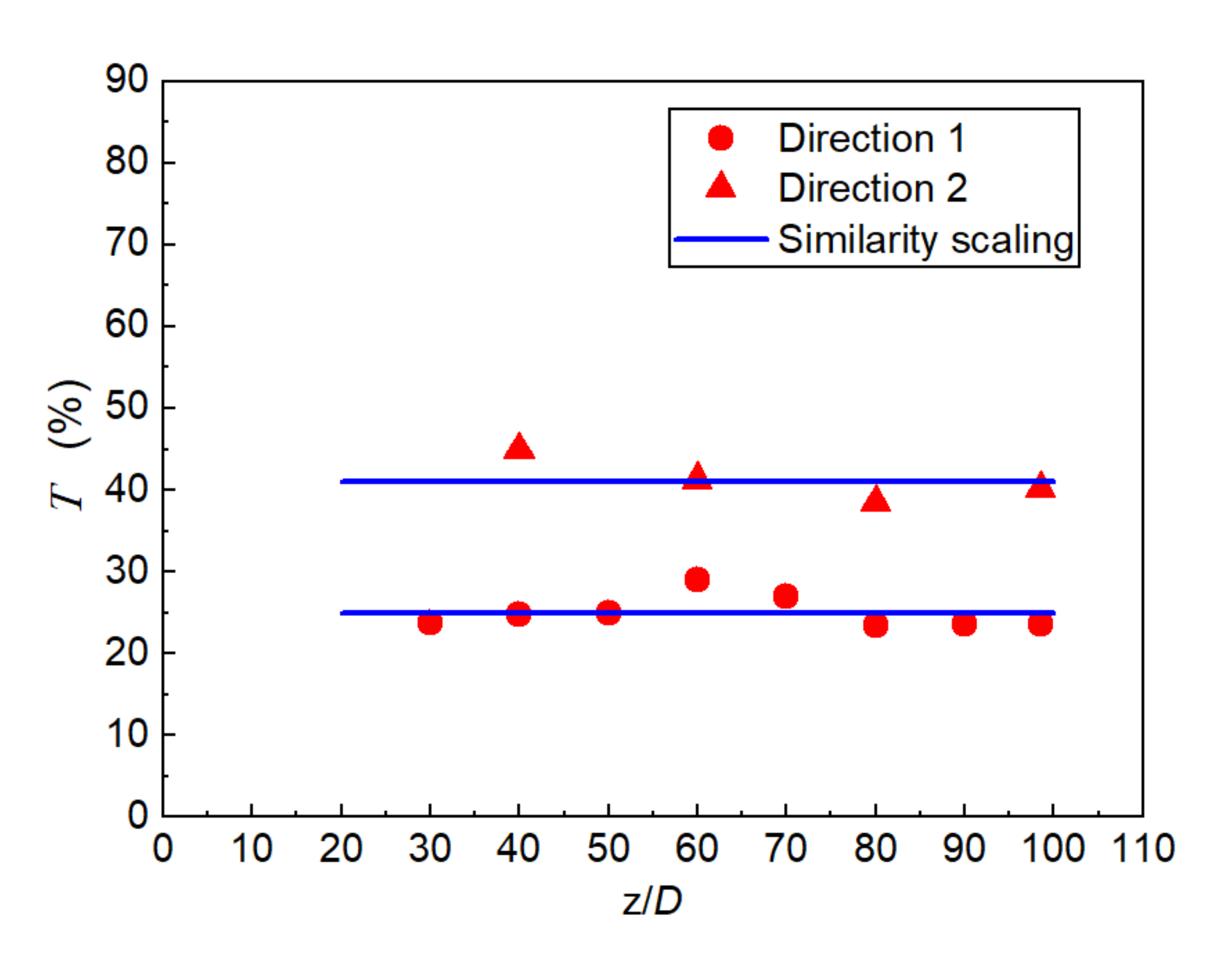}
    \caption{Turbulence intensity of the axial velocity component along Direction 1 ($25\,\%$) and Direction 2 ($41\,\%$), respectively. The measurements have been compared to constant values, in accordance with the expected similarity scaling.}
    \label{fig:8}
\end{figure}

Figure~\ref{fig:8} illustrates the turbulence intensity of the axial velocity component along Directions 1 and 2. Direction 2 is of particular interest, since it indicates the usefulness of a spherical coordinate system as originally proposed by~\cite{AzurThesis}. The turbulence intensity is, as highlighted by the fitted constant values, to within the experimental accuracy constant along radial lines and thus adhere to the similarity scaling as expected from the simple model.

\subsection{Dynamic moments}\label{sec:DynMom}

In this section we consider the measured second and third order dynamic moments. We shall investigate how these statistical functions scale along the jet axis and along the direction transverse to the jet axis and compare to the scaling expected from the simple model in~\cite{JetSimPart1}. Again, these self-similarity predictions from the simple model are denoted ``Similarity scaling''.  

We shall compute the dynamic spatial moments based on the convection record~\cite{buchhave2017measurement}. As the convection record for a stationary flow can be considered a spatial, homogeneous record measured at one point in the jet, we can directly compute the local, single point spatial power spectrum.

\subsubsection{Power spectral density}

\begin{figure}[t]
    \begin{minipage}{\linewidth}
    \centering
    \includegraphics[width=0.9 \linewidth]{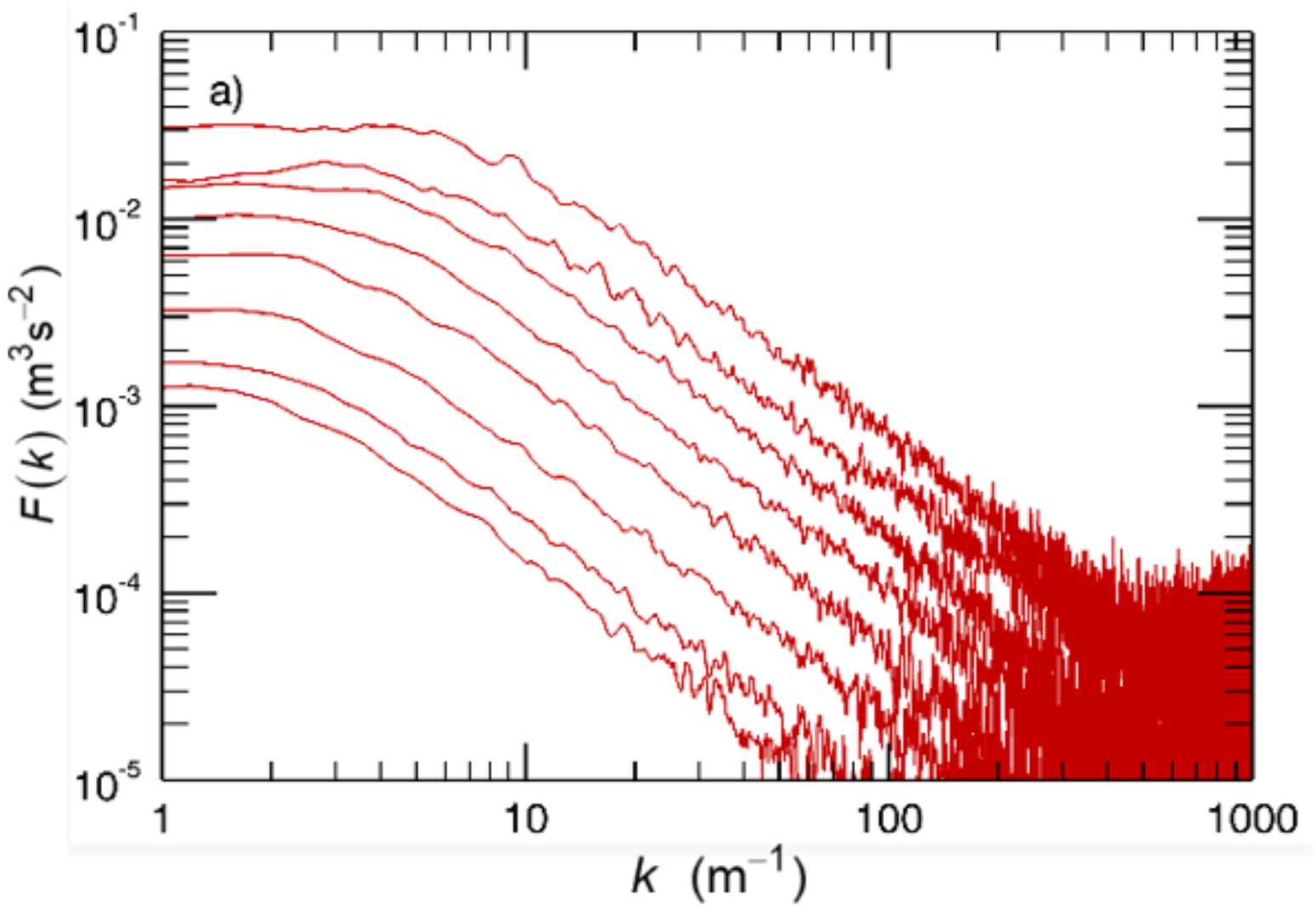}\\
    {\textit{(a)} Raw PSD spectra.}\vspace{0.3cm}
    \end{minipage}
    \begin{minipage}{\linewidth}
    \centering
    \includegraphics[width=0.9 \linewidth]{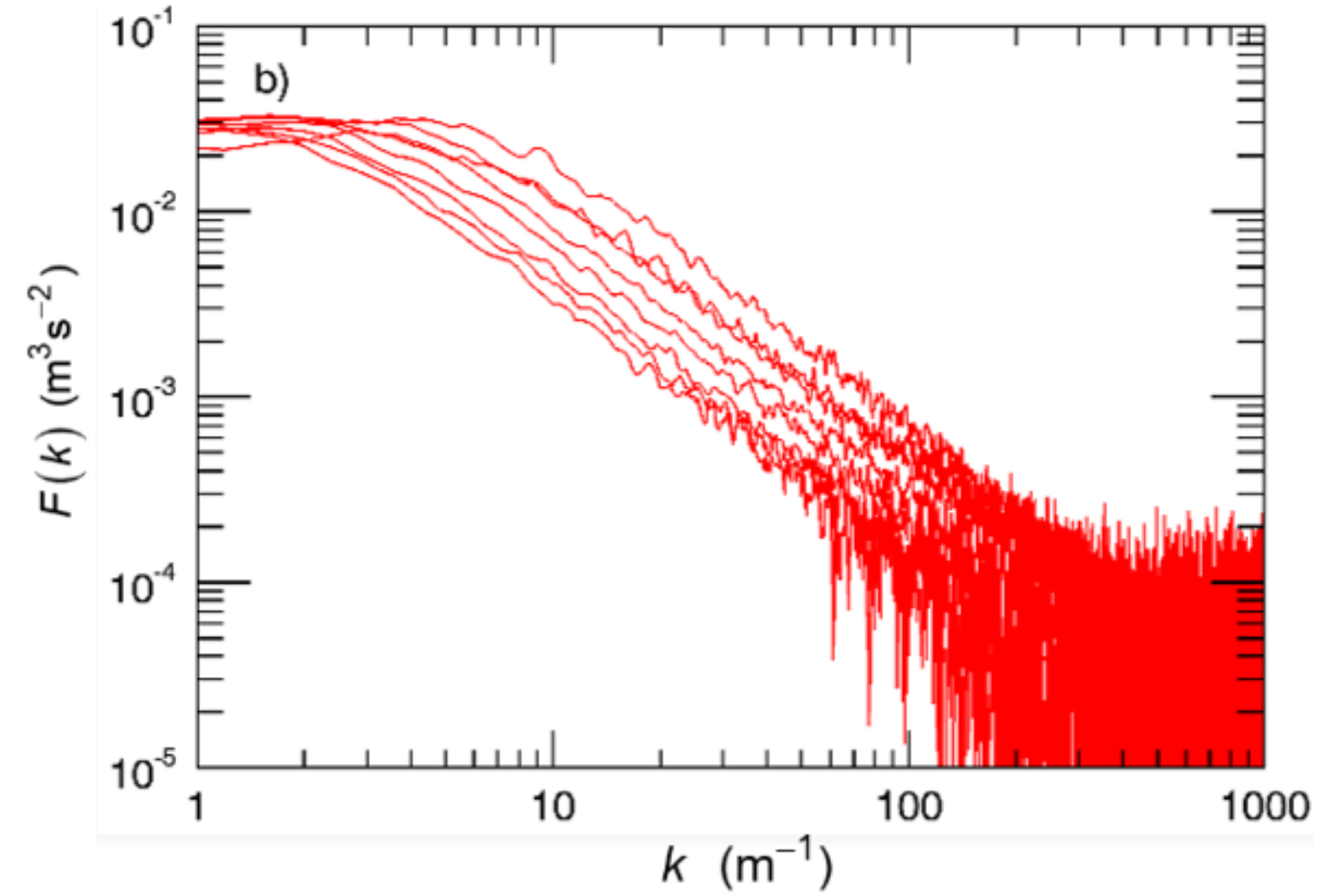}\\
    {\textit{(b)} Scaling of the ordinate.}\vspace{0.3cm}
    \end{minipage}
        \begin{minipage}{\linewidth}
    \centering
    \includegraphics[width=0.95 \linewidth]{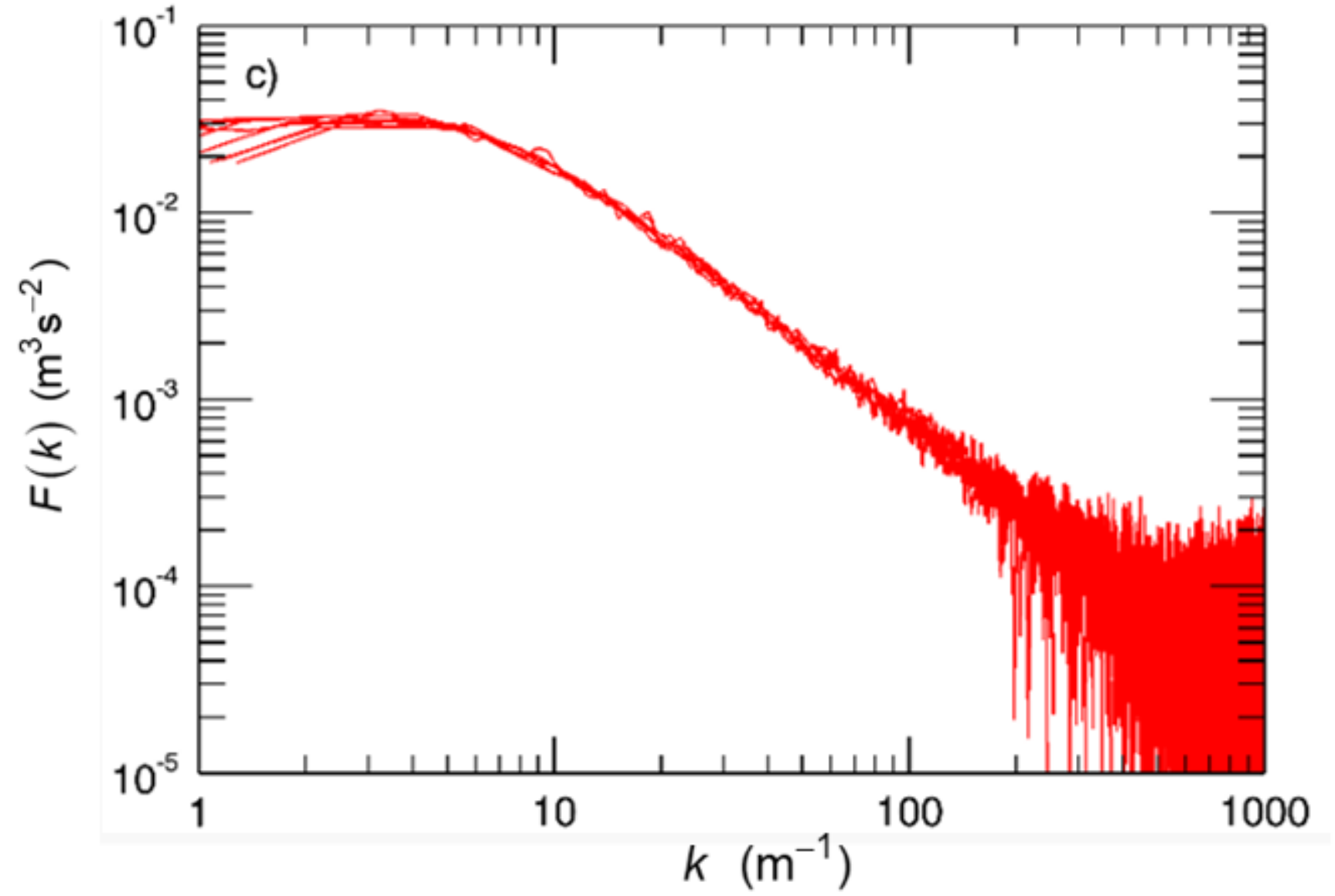}\\
    {\textit{(c)} Scaling of both ordinate and abscissa.} \vspace{0.3cm}
    \end{minipage}
    \caption{Downstream developments of spatial power spectral densities at along the jet centerline. All spectra, right to left: $z/D =30 \, \rightarrow\, 100$.}
    \label{fig:9}
\end{figure}

Because of the finite record length, the computed power spectral density, $F(k)$, is the convolution of the true spectrum, $F(k)_{true}$, and the box car spectral window (a sinc-squared function):
$$
F(k)=F(k)_{true} \otimes L_m \, \mathrm{sinc}^2 \left ( \frac{1}{2}k L_m \right )
$$
where $L_m$ is the length of the $m$'th spatial record. In order to compute the correct power spectral density, we must correct the computed power spectrum with the weight of the spectral window, also called the equivalent noise band width (ENBW). However, the lengths of the spatial records vary because of the fluctuating velocity, and we must compute the ENBW for each record and correct all the records individually. Random sampling causes a low, flat noise level in each PSD, which we remove in the final plot. Figure~\ref{fig:9}$a$ shows the measured PSD along the jet centerline where the downstream development follows the shift in spectra from right to left, $z/D = 30 \, \rightarrow \, 100$.

To test these ideas, we have extracted the values of the factors needed to scale all the measured PSDs to collapse with the PSD at $30D$, as shown in Figure~\ref{fig:9}. Figure~\ref{fig:9}$a$ shows the raw measured spectra. Figure~\ref{fig:9}$b$ shows the result of multiplying the ordinate of the downstream measured PSDs by a factor $F_{ord}$, to the same ordinate value as the PSD at $30D$. Figure~\ref{fig:9}$c$ shows similarly the result of also scaling the downstream PSDs along the abscissa by a factor, $F_{abs}$ to collapse with the PSD at $30D$ along the $k$-axis.

\begin{figure}[t]
    \begin{minipage}{\linewidth}
    \centering
    \includegraphics[width=\linewidth]{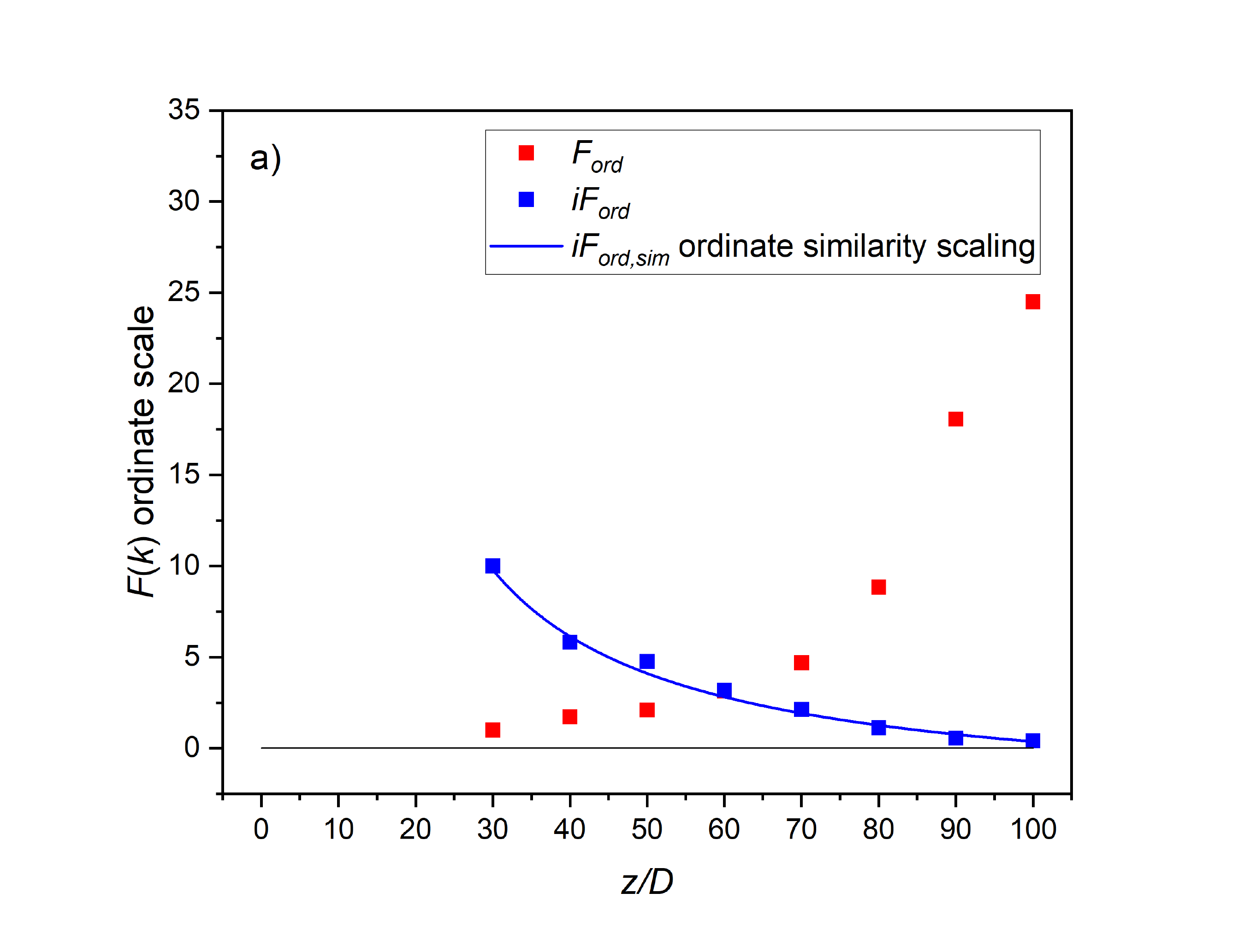}
    {\textit{(a)} Ordinate scaling of $F(k)$.}
    \end{minipage}
    \begin{minipage}{\linewidth}
    \centering
    \includegraphics[width=\linewidth]{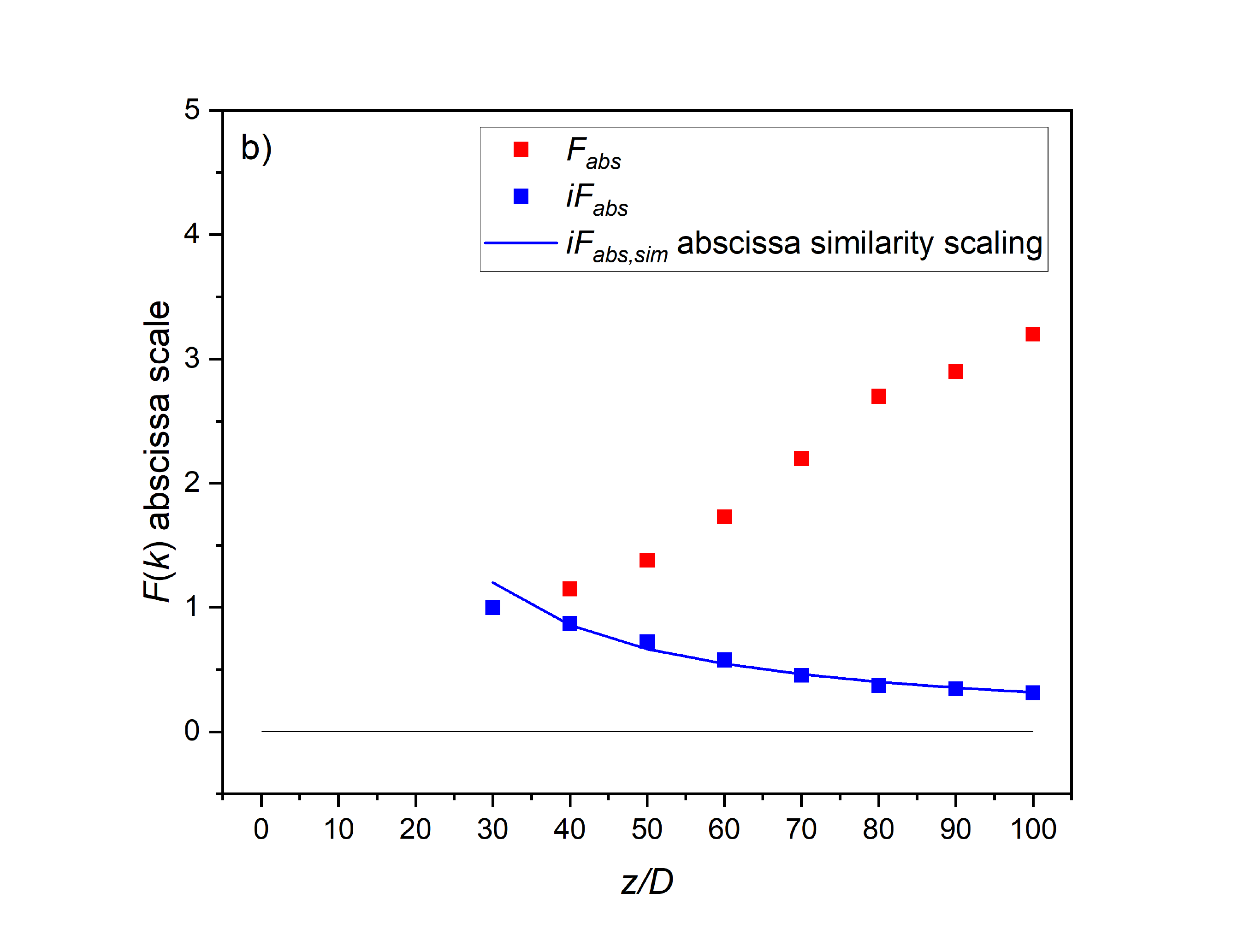}
    {\textit{(b)} Abscissa scaling of $F(k)$.}
    \end{minipage}
    \caption{a) Scaling factor to collapse PSD along ordinate axis. Red squares: Measured $F_{ord}$. Blue squares: Inverted ordinate scaling factor, $iF_{ord}$. Blue line: Expected PSD similarity scaling, $iF_{ord,sim}$. b) Scaling factor to collapse PSD along $k$-axis. Red squares: $F_{abs}$. Blue squares: Inverted abscissa factor, $iF_{abs}$. Blue line: Expected abscissa similarity scaling, $iF_{abs,sim}$.}
    \label{fig:10}
\end{figure}

Figure~\ref{fig:10}$a$ shows the measured values of $F_{ord}$ as well as its inverse $iF_{ord}$. Also shown is the ordinate scaling factor expected from the simple model, namely an inverse first order dependence on the distance to the virtual origin, $\left ( z-z_0 \right )$: $iF_{ord,sim} = 320 \cdot \left ( z-z_0 \right )^{-1} - 3.0$ with $z_0/D=5$. 

Figure~\ref{fig:10}$b$ shows correspondingly the measured values of $F_{abs}$ as well as its inverse $iF_{abs}$ needed to collapse the PSDs along the abscissa. Also shown is the fitting function following the expected form from the simple model: $iF_{abs,sim} = 30 \cdot \left ( z-z_0 \right )^{-1}$ with $z_0/D=5$.

These results support that the power spectral density scales along the jet center axis according to a single, geometric scaling factor that equals, to within experimental uncertainty, the distance from a single virtual origin approximately located five jet exit diameters downstream of the jet orifice. 

As illustrated in Figures~\ref{fig:9} and~\ref{fig:10}, these first order scaling functions convincingly collapse all the downstream spectra to a single PSD curve, in this case the PSD at $30D$. In other words, this scaling of the PSD appears to confirm that the scaling of the free, round jet, in agreement with the predictions of~\cite{JetSimPart1}, can be traced to basic physical conservation laws and does not need to rely on ad hoc assumptions about the existence of similarity transformations or on previous experimental data. We also note that we see a scaling from the same value of the virtual origin as we saw in the first order statistical functions such as mean velocity and the jet width.

\begin{figure}
    \begin{minipage}{\linewidth}
    \centering
    \includegraphics[width=\linewidth]{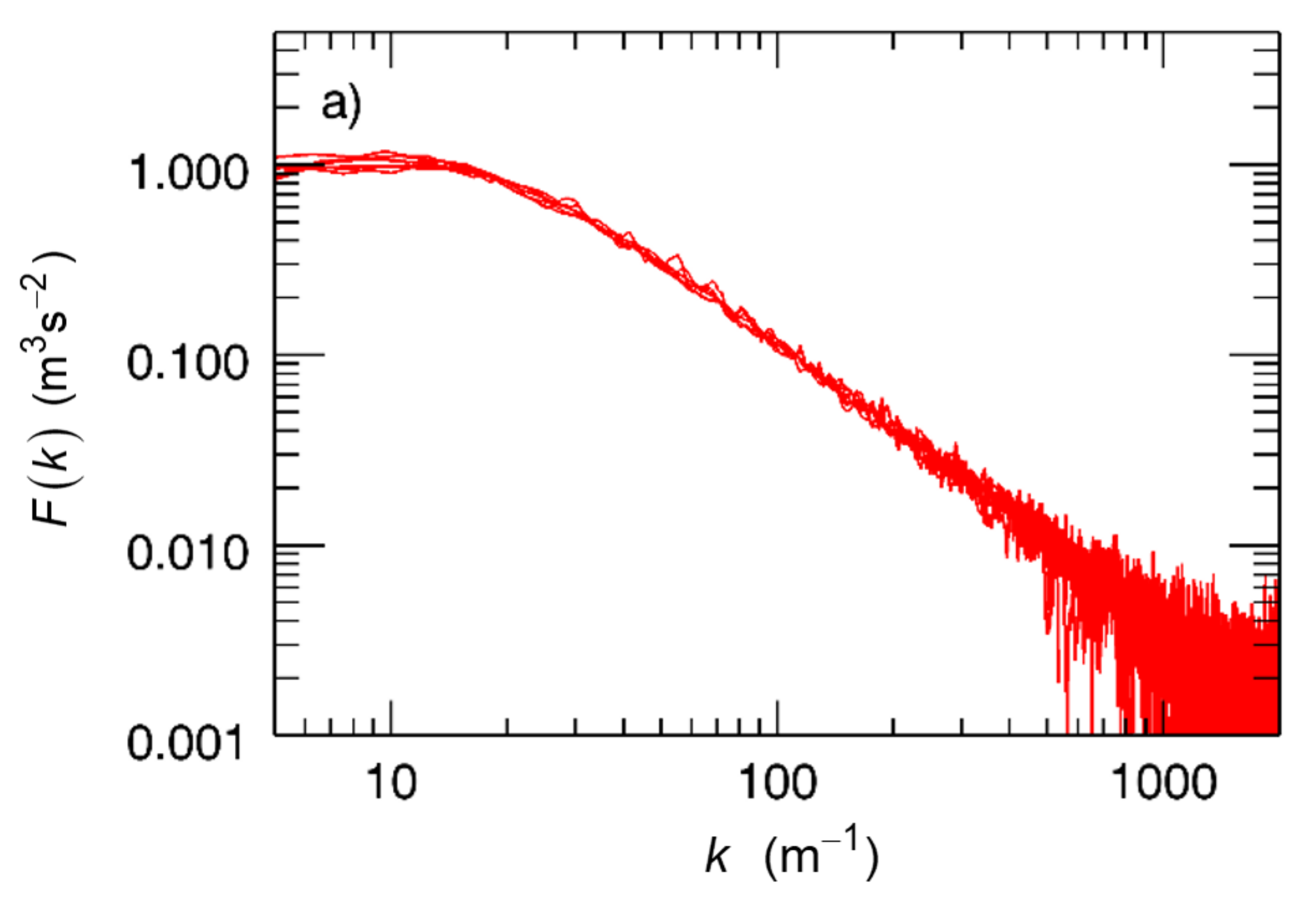}\\
    {\textit{(a)} Direction 1}\vspace{0.3cm}
    \end{minipage}
    \begin{minipage}{\linewidth}
    \centering
    \includegraphics[width=\linewidth]{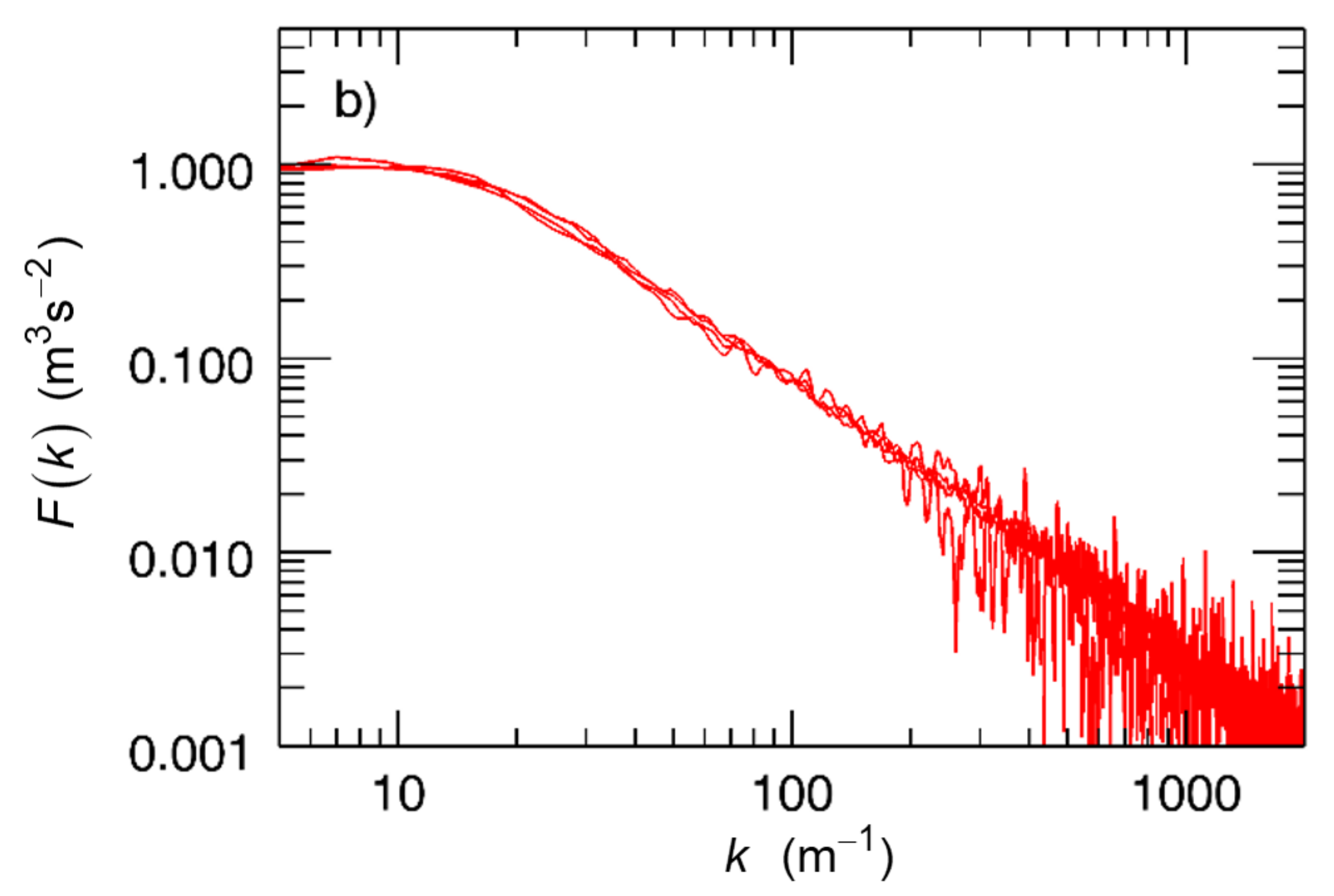}\\
    {\textit{(a)} Direction 2}\vspace{0.3cm}
    \end{minipage}
    \caption{Normalized PSDs along a) Direction 1 and b) Direction 2.}
    \label{fig:11}
\end{figure}

To further highlight the similarity of the PSDs along the natural jet development directions, we present PSD plots along Direction 1 (Figure~\ref{fig:11}$a$) and Direction 2 (Figure~\ref{fig:11}$b$) where the maximum values of each PSD has been normalized to unity.

\begin{figure}
    \begin{minipage}{\linewidth}
    \centering
    \includegraphics[width=\linewidth]{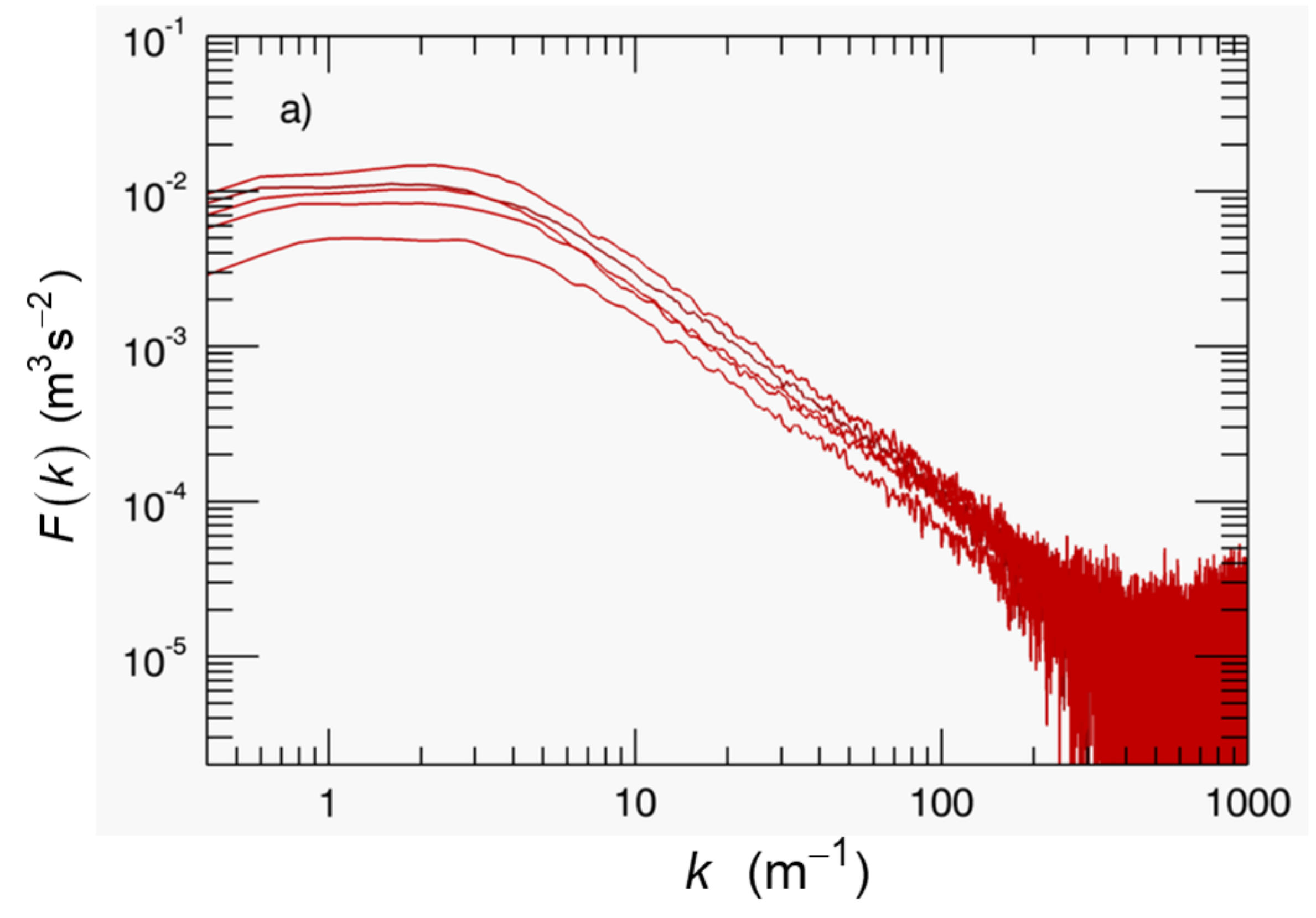}\vspace{0.5cm}
    \end{minipage}
    \begin{minipage}{\linewidth}
    \centering
    \includegraphics[width=\linewidth]{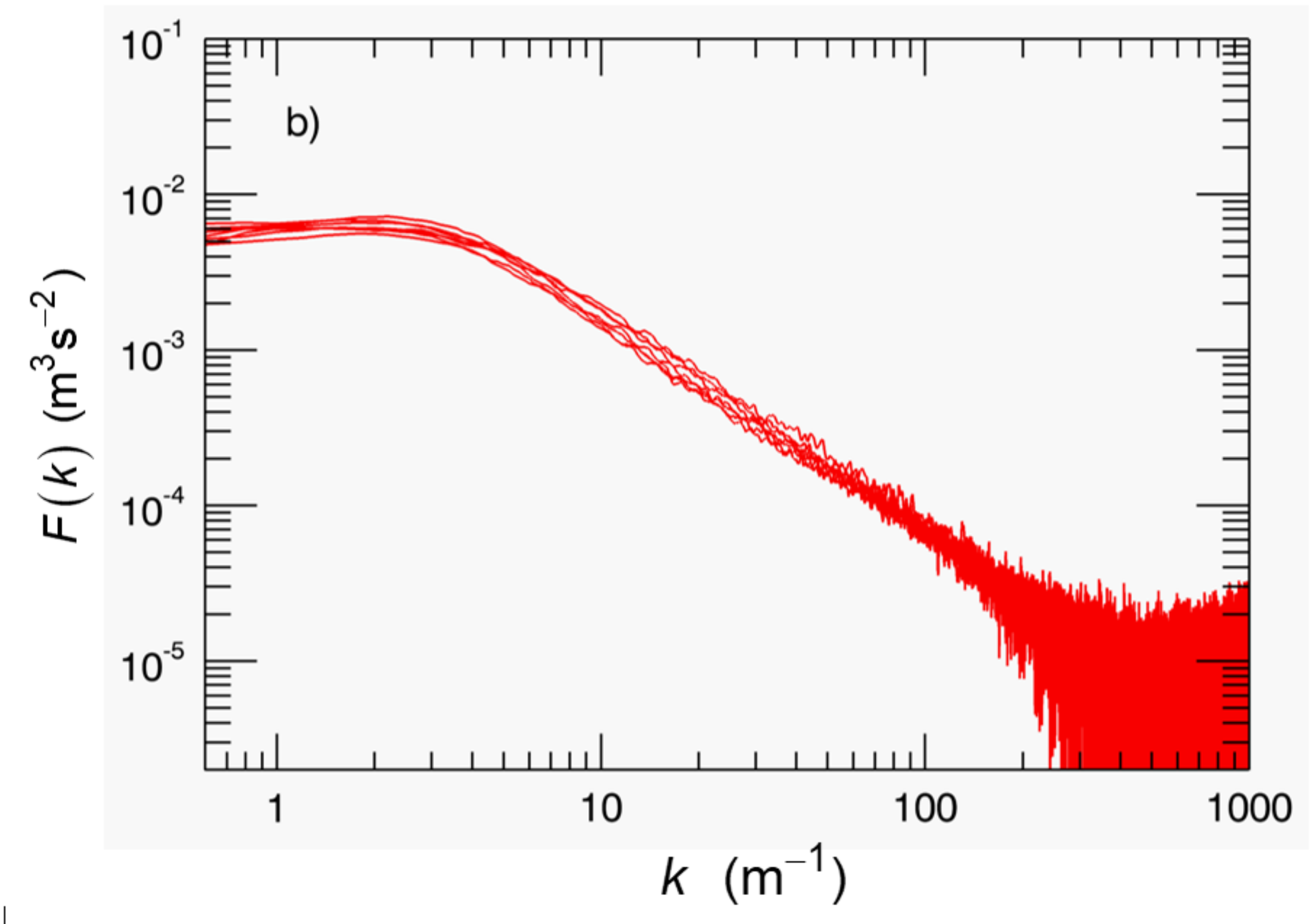}
    \end{minipage}
    \caption{Raw PSD spectra measured across the jet at $z/D=60$ (increasing radial distance from right to left). b) PSDs across the jet at $z/D=60$, normalized by the variance at $z/D=60$ at the centerline.}
    \label{fig:12}
\end{figure}

The PSD variations at a fixed distance $z$ from the origin is of great interest, as it describes the distribution of the kinetic energy fluctuations across the jet. In Figure~\ref{fig:12}$a$, we plot the raw PSDs at $z/D=60$ for the measured radial positions. In Figure~\ref{fig:12}$b$, we have normalized the ordinate of the measured PSDs by adjusting the total power to be equal to that at the jet centerline at $z/D=60$ in order to better be able to compare the forms of the spectra. Obviously, the spatial fluctuations are similar across the jet at a fixed distance from the origin, indicating an efficient mixing of spatial scales. 

\begin{figure}
    \begin{minipage}{\linewidth}
    \centering
    \includegraphics[width=\linewidth]{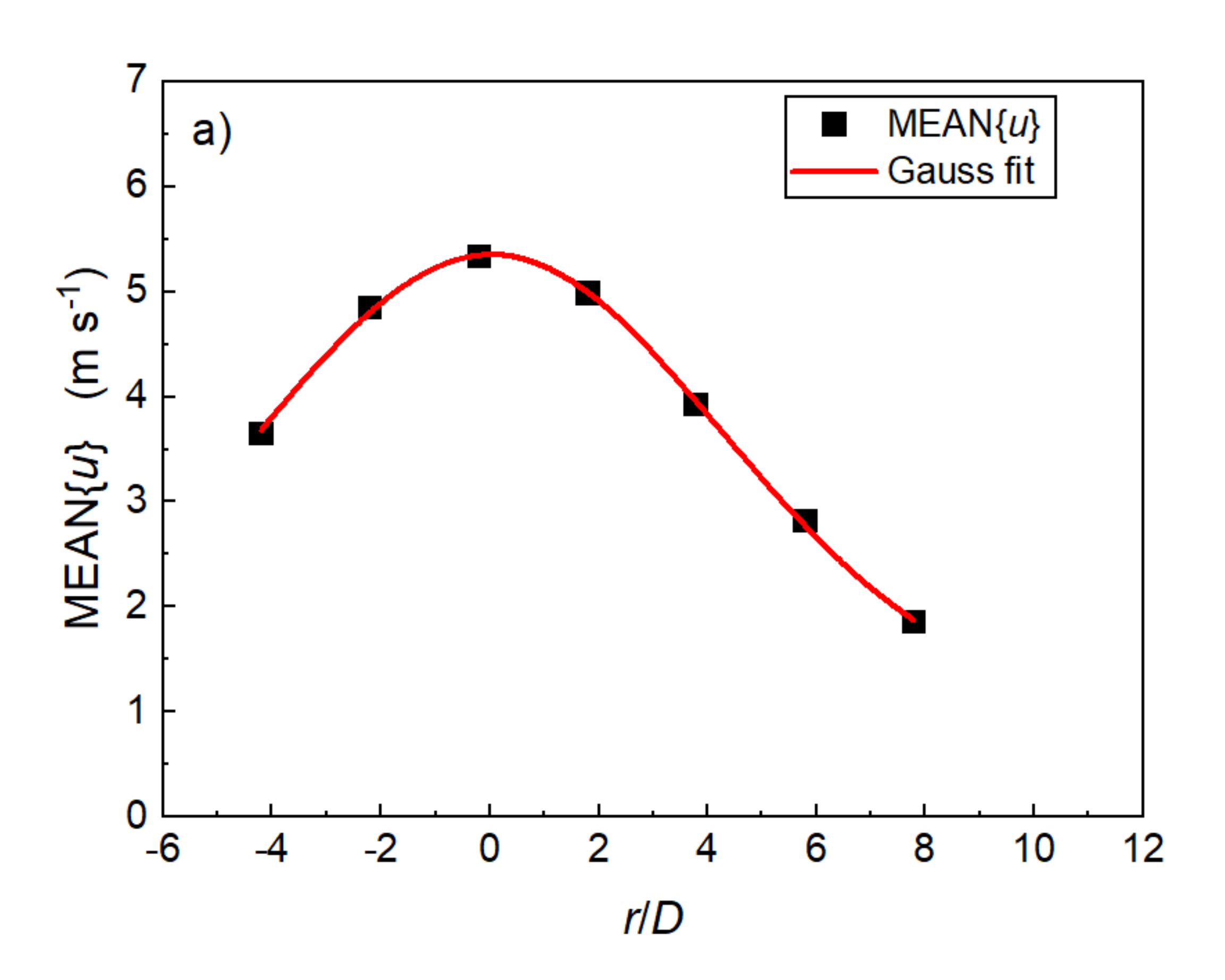}
    \end{minipage}
    \begin{minipage}{\linewidth}
    \centering
    \includegraphics[width=\linewidth]{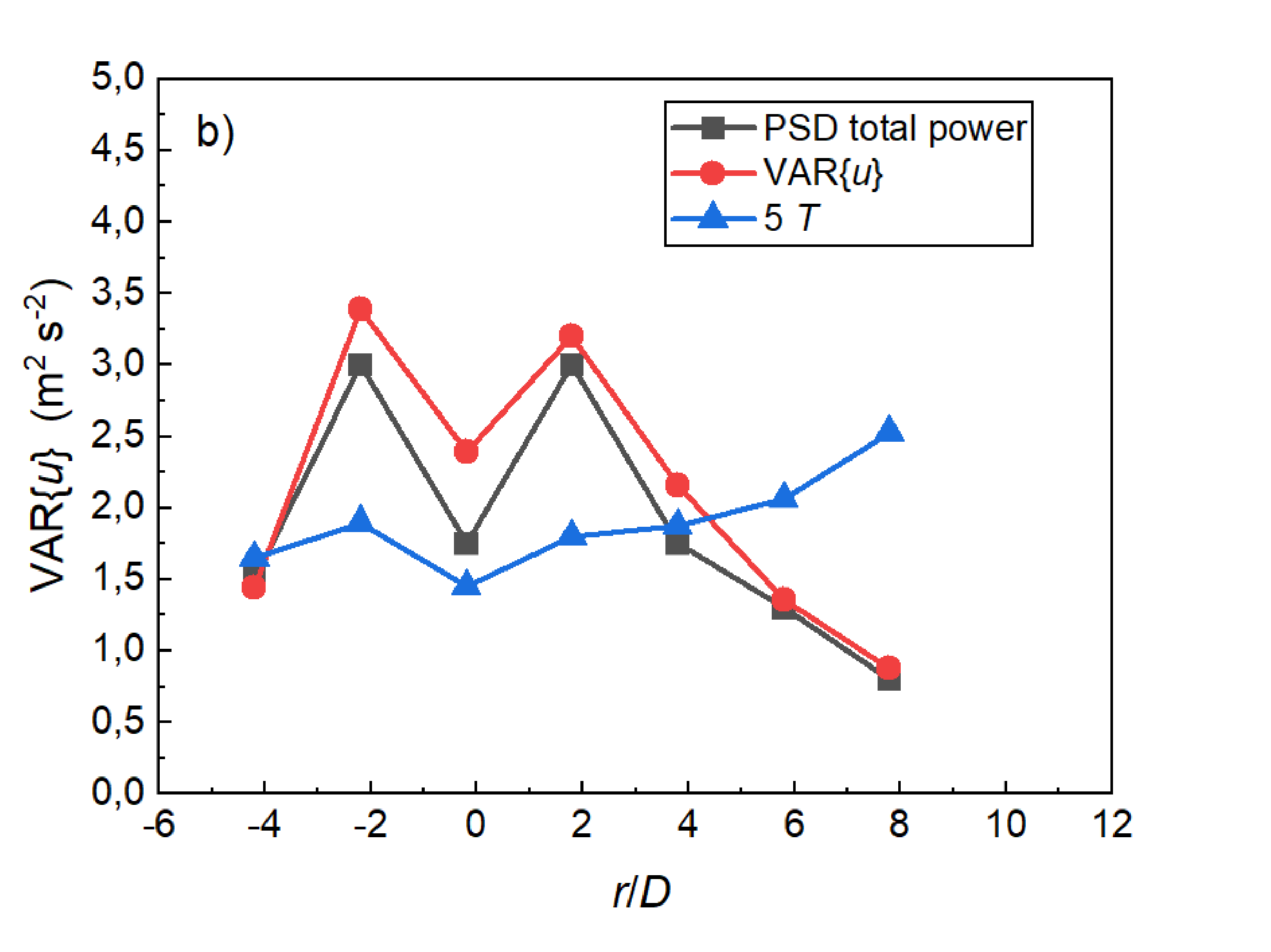}
    \end{minipage}
    \caption{a) Mean streamwise velocity profile along the radial direction at $z/D=60$. The red curve is a Gaussian fit. b) Comparison of the total integrated power from the PSDs (black squares) and the corresponding variance (red dots) across the radial direction of the jet. Also shown for comparison is the local turbulence intensity (blue triangles, values multiplied by 5). Points connected to aid the eye.}
    \label{fig:13}
\end{figure}

Figure~\ref{fig:13}$a$ shows the mean streamwise velocity profile in the radial direction at $z/D=60$ along with a Gaussian fit to the mean velocity. In Figure~\ref{fig:13}$b$, corresponding radial profiles of the total power (integral) of the PSDs and the respective profile of the measured variance are shown. The total power in the PSDs should, by definition, equal to the local variance. The difference can be attributed to the self-products in the computation of the variance. The noise from the self-products results from high frequency noise unrelated to the measured turbulence. This noise has been subtracted from the PSDs and is thus not present in the profile of the total power of the PSDs. Figure~\ref{fig:12}$b$ also shows the local turbulence intensity (multiplied by $5$) in the transverse direction, for reference. From this figure, we see that the spatial velocity scales are efficiently mixed and has the same shape across the jet, and that the fluctuations relative to the mean velocity are greater at the edge of the jet. 

\begin{figure}[t]
    \begin{minipage}{\linewidth}
    \centering
    \includegraphics[width=0.85 \linewidth]{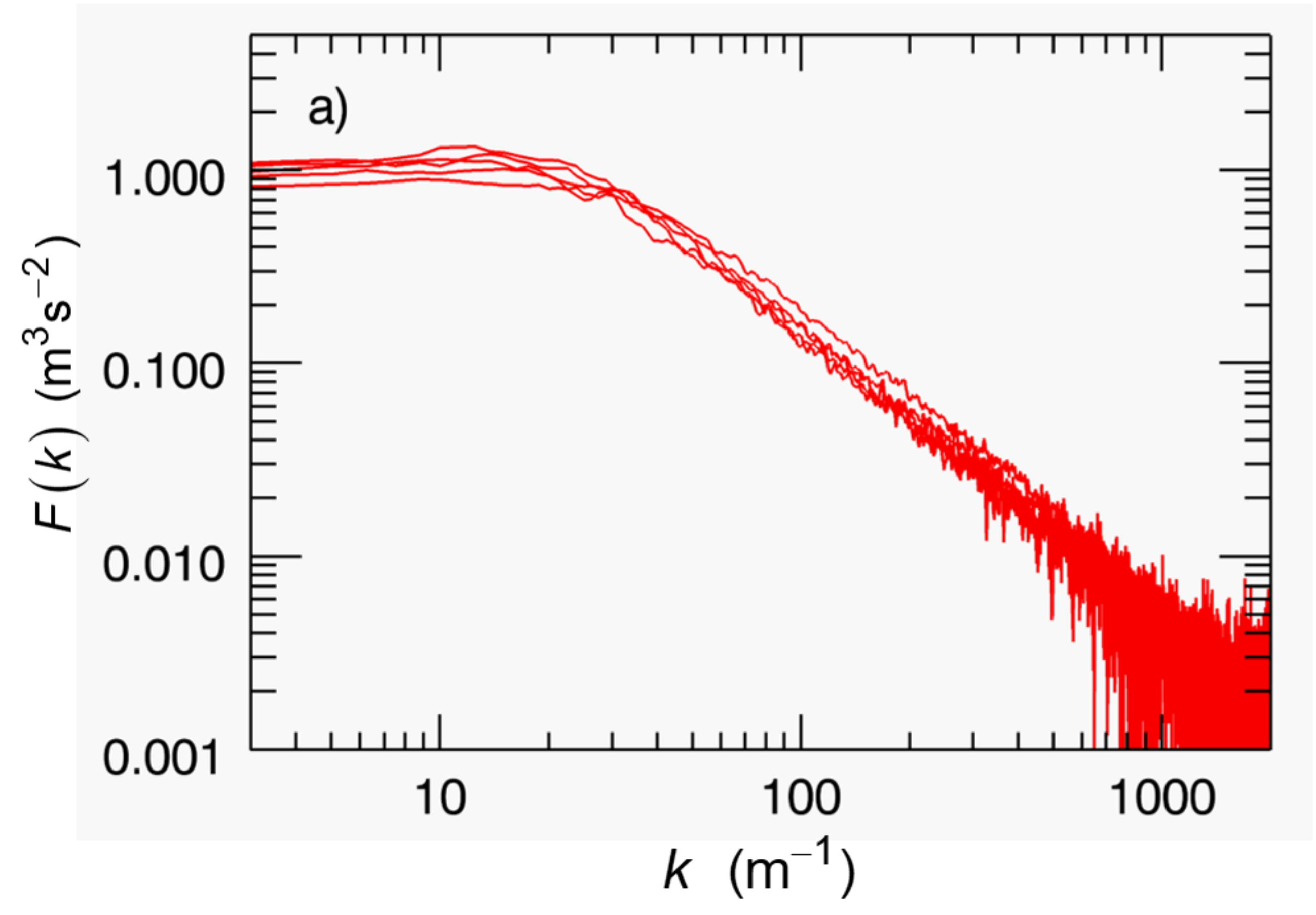}\\
    {\textit{(a)} $z/D = 30$}\vspace{0.3cm}
    \end{minipage}
    \begin{minipage}{\linewidth}
    \centering
    \includegraphics[width=0.85 \linewidth]{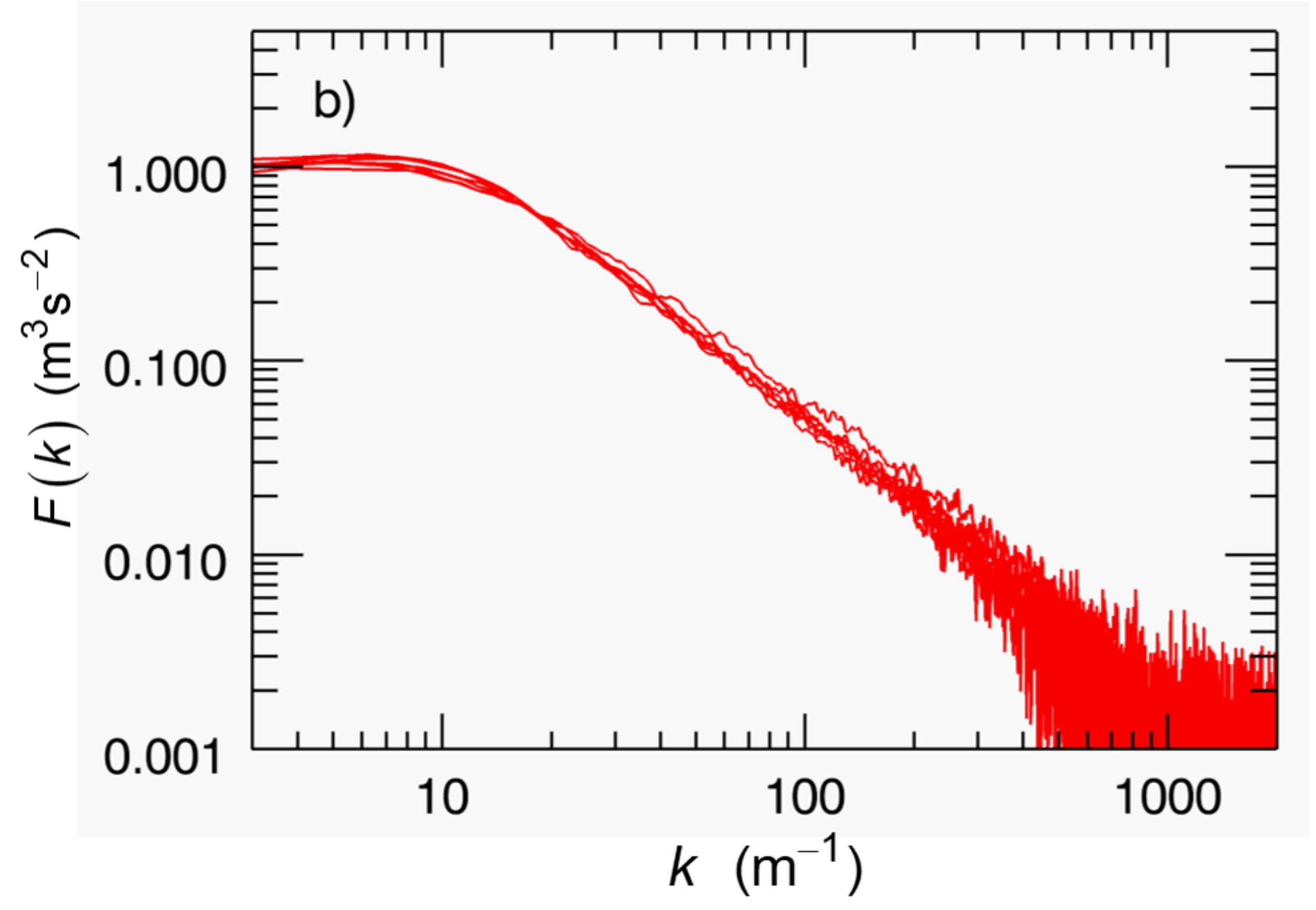}\\
    {\textit{(b)} $z/D = 60$}\vspace{0.3cm}
    \end{minipage}
    \begin{minipage}{\linewidth}
    \centering
    \includegraphics[width=0.85 \linewidth]{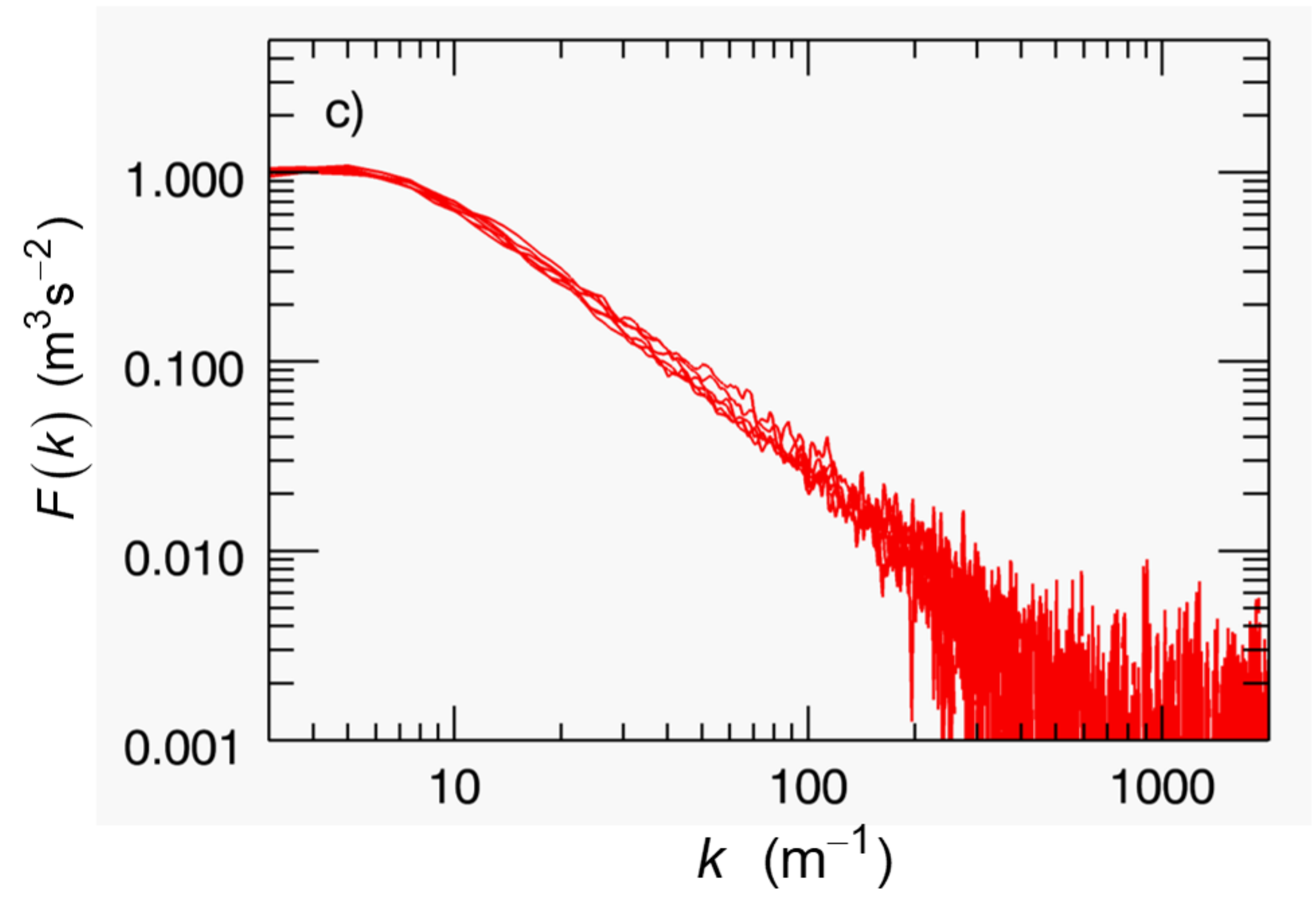}\\
    {\textit{(c)} $z/D = 90$}\vspace{0.3cm}
    \end{minipage}
    \caption{Transverse development of the normalized PSDs at the respective measured cross sectional positions at axial distances: a) $z/D=30$, b) $z/D=60$ and c) $z/D=90$. }
    \label{fig:14}
\end{figure}

Figures~\ref{fig:14}$a-c$ show the normalized PSDs collapsing in cross sections at $z/D=30$, $60$ and $90$, respectively. 

\subsubsection{Covariance function}

We compute the spatial autocovariance function (ACF) using the slotting method (c.f.~\cite{AnnurevBGL}) by sorting products of velocities at the actually occurring random arrival convection record points into equidistant displacement slots (the so-called slotted autocovariance function, SACF). 

Again, we emphasize that the mapping from time lags to spatial displacements that we obtain by using the residence times means that we can compute the single point covariance of a homogeneous record as long as the flow is stationary,~\cite{buchhave2017measurement}. This means that we fulfill the prerequisite for identical joint velocity probability distributions (homogeneity) of the two velocities (points) in the covariance product required for a meaningful statistical function,~\cite{monin2007statistical} or~\cite{frisch1995turbulence}. This is in contrast to many two-point covariance measurements in jets, where the joint probability distributions cannot be considered identical as the jet is not a homogeneous flow along the axial direction.

\begin{figure}[t]
    \begin{minipage}{\linewidth}
    \centering
    \includegraphics[width=0.85 \linewidth]{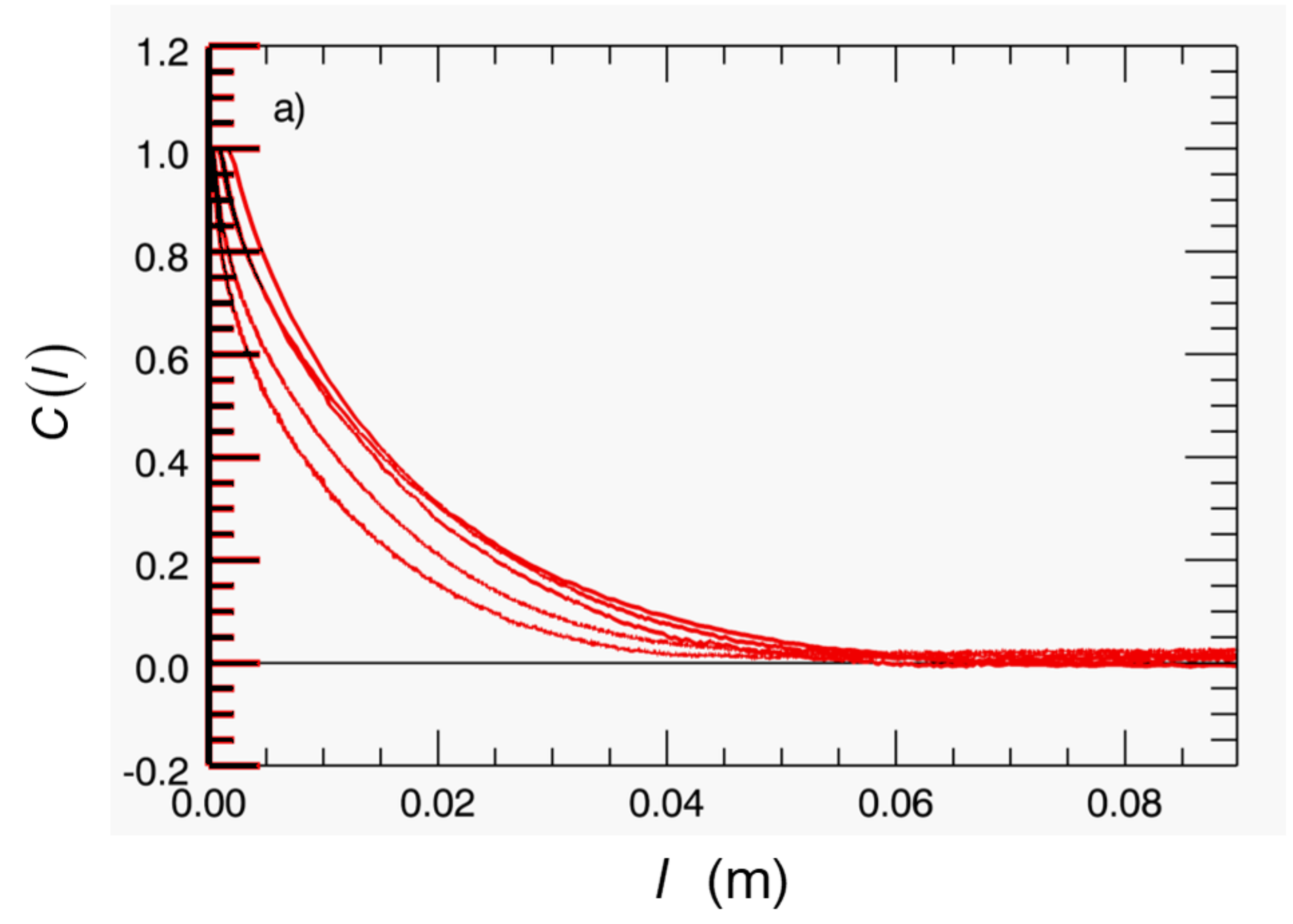}\\
    {\textit{(a)} $z/D = 30$}\vspace{0.3cm}
    \end{minipage}
    \begin{minipage}{\linewidth}
    \centering
    \includegraphics[width=0.85 \linewidth]{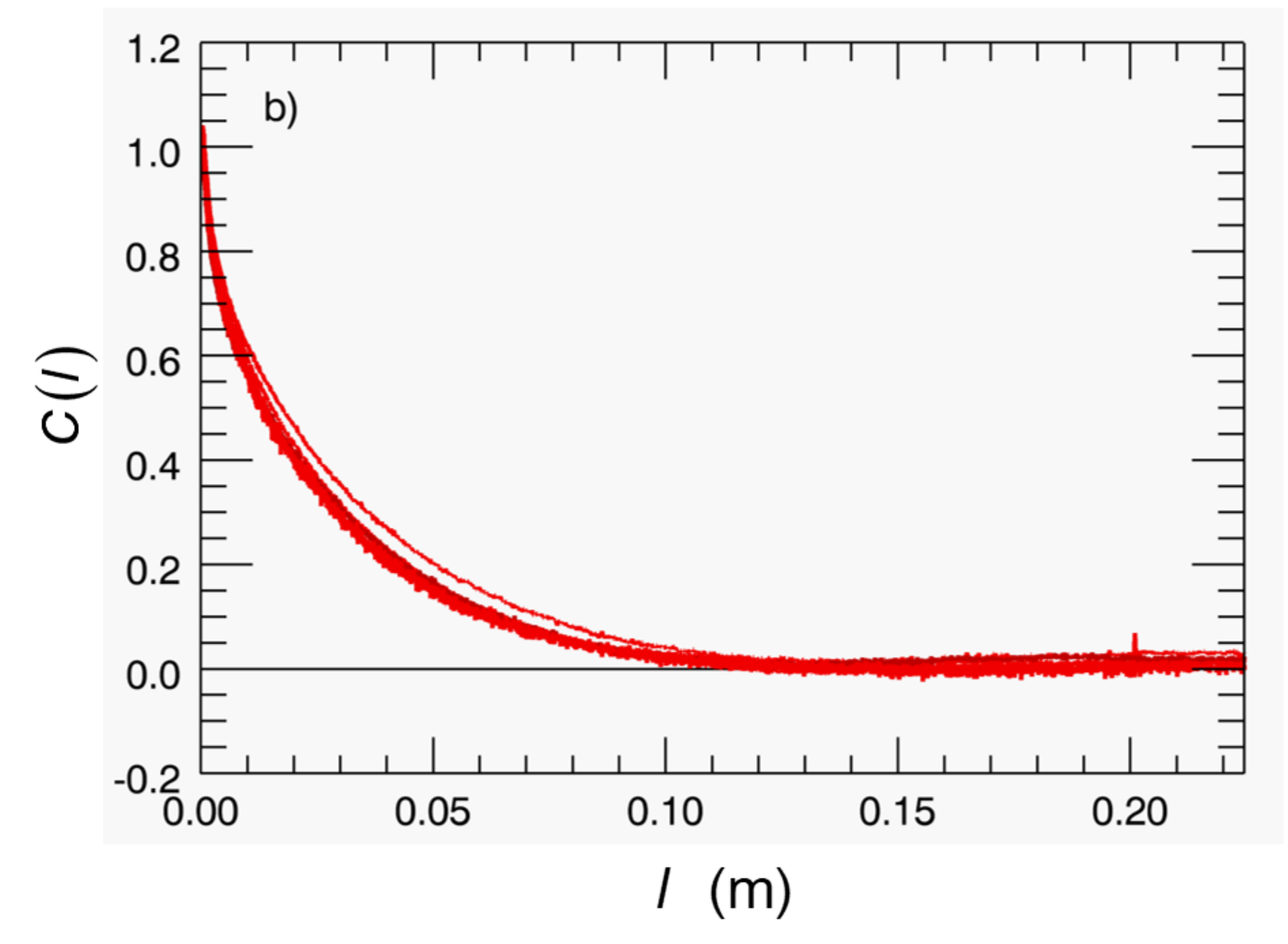}\\
    {\textit{(b)} $z/D = 60$}\vspace{0.3cm}
    \end{minipage}
    \begin{minipage}{\linewidth}
    \centering
    \includegraphics[width=0.85 \linewidth]{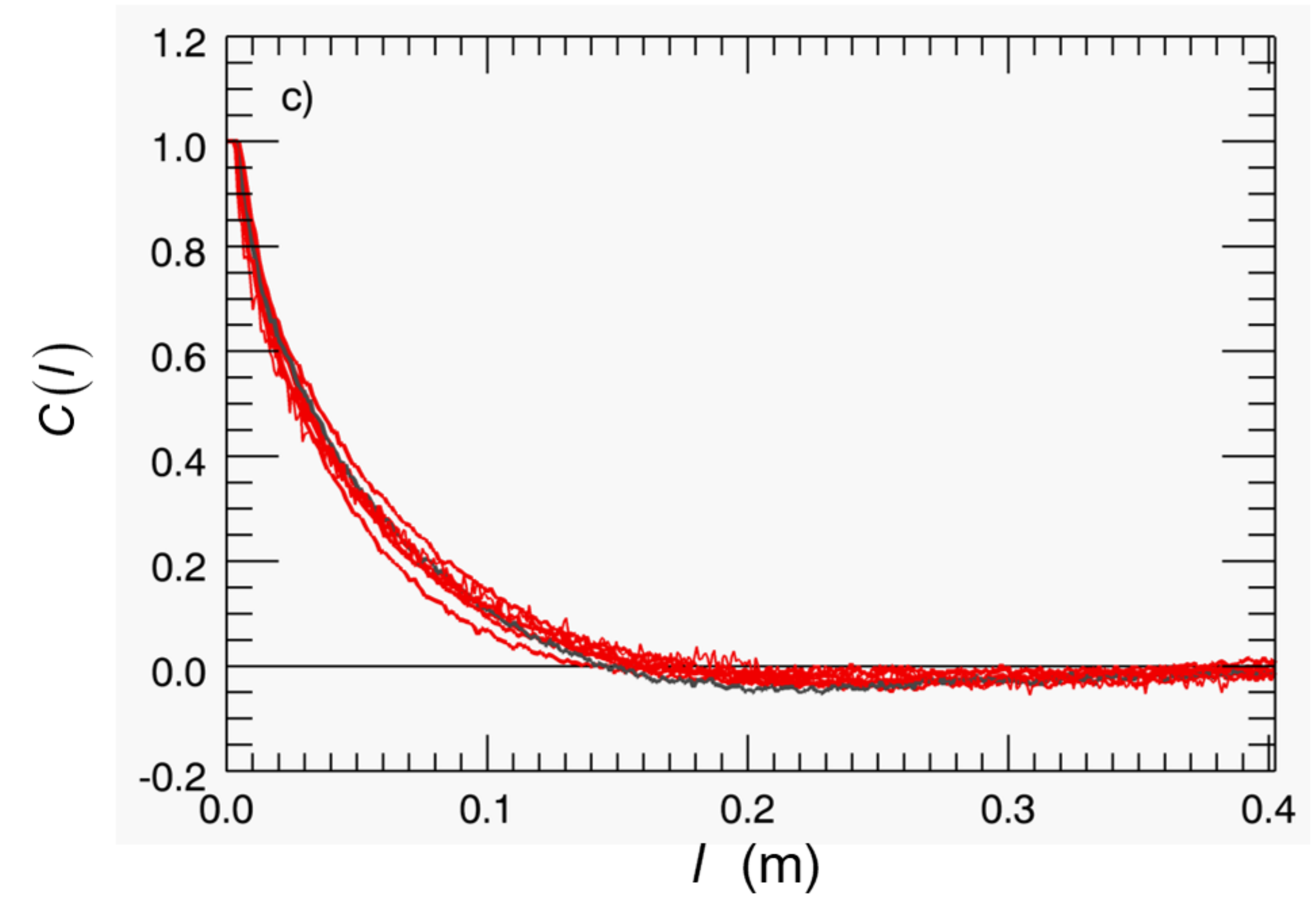}\\
    {\textit{(c)} $z/D = 90$}\vspace{0.3cm}
    \end{minipage}
    \caption{Transverse measurements of the ACF at the respective measured cross sectional positions at axial distances: a) $z/D=30$, b) $z/D=60$ and c) $z/D=90$ as a function of spatial lag, $l$.}
    \label{fig:15}
\end{figure}

Figures~\ref{fig:15}$a-c$ show the autocorrelation function, $C(l)$, computed at cross sectional scans at $z/D=30$, $60$ and $90$, respectively. As before, we observe that the structures are the same across the jet at different downstream positions. 

In all the computed ACFs, a spike, which is approximately 30\% higher than that at the following lag, occurs at zero lag. We attribute the added contribution to $C(l=0)$ to self-products from high frequency noise not related to the turbulence. This is the same self-product effect that causes the variance to be higher than the total power computed from the PSD. In Figure~\ref{fig:15}, we have extrapolated the ACF from the first couple of points back to zero lag with a second order polynomial with horizontal tangent.

\subsubsection{Structure functions}

As described in section~\ref{subsec:signdatproc}, the second and third order structure functions are computed from the measured axial velocity records at the random spatial separations or displacements, $s_n-s_{n'}$, in the convection record, raised to the second power and third power, respectively, and averaged over the record. The displacements are sorted into equal size bins or slots. The computations are repeated for each record and then averaged over all records. 

\begin{figure}[t]
    \begin{minipage}{\linewidth}
    \centering
    \includegraphics[width=0.85\linewidth]{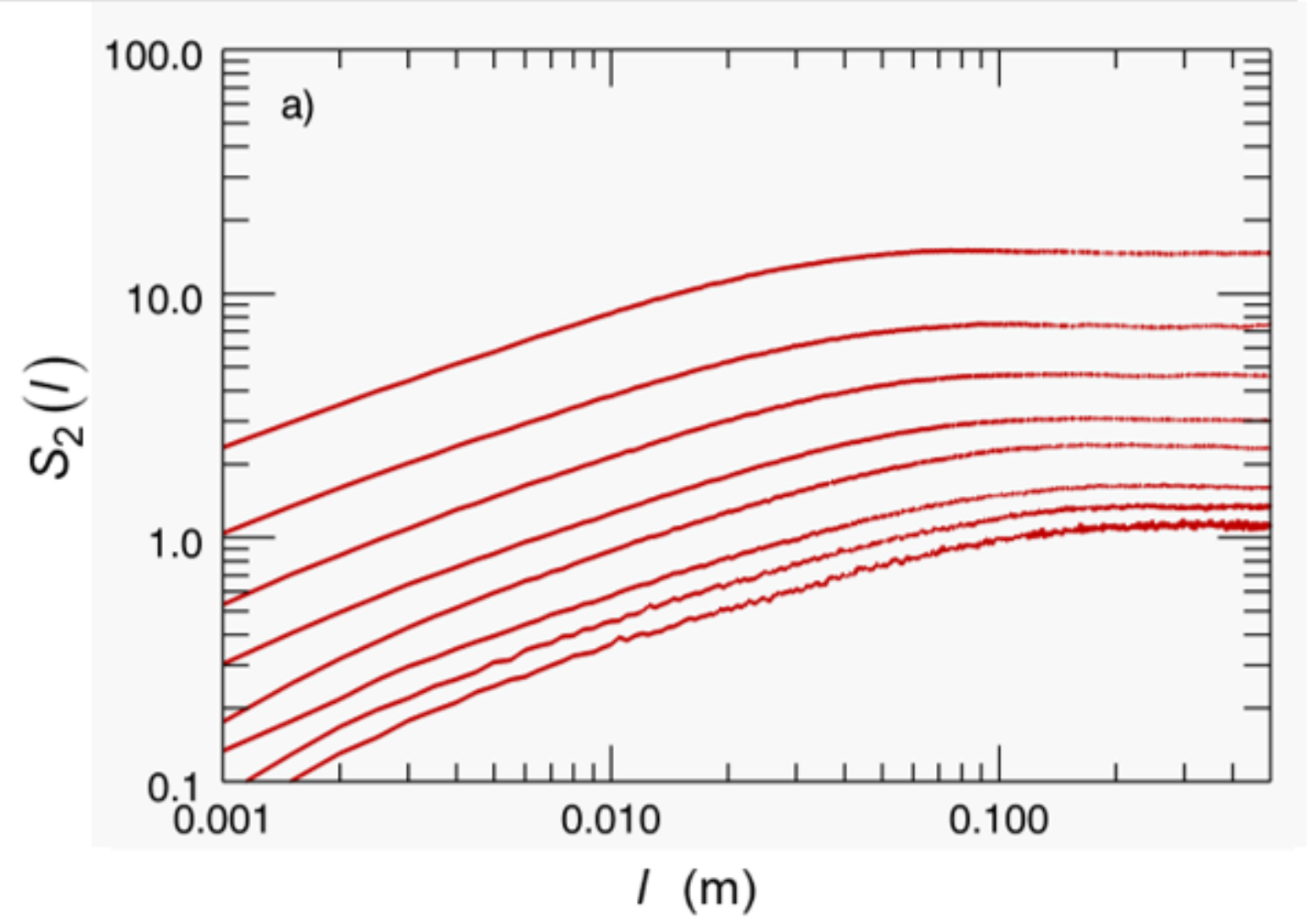}\vspace{0.5cm}
    \end{minipage}
    \begin{minipage}{\linewidth}
    \centering
    \includegraphics[width=0.85\linewidth]{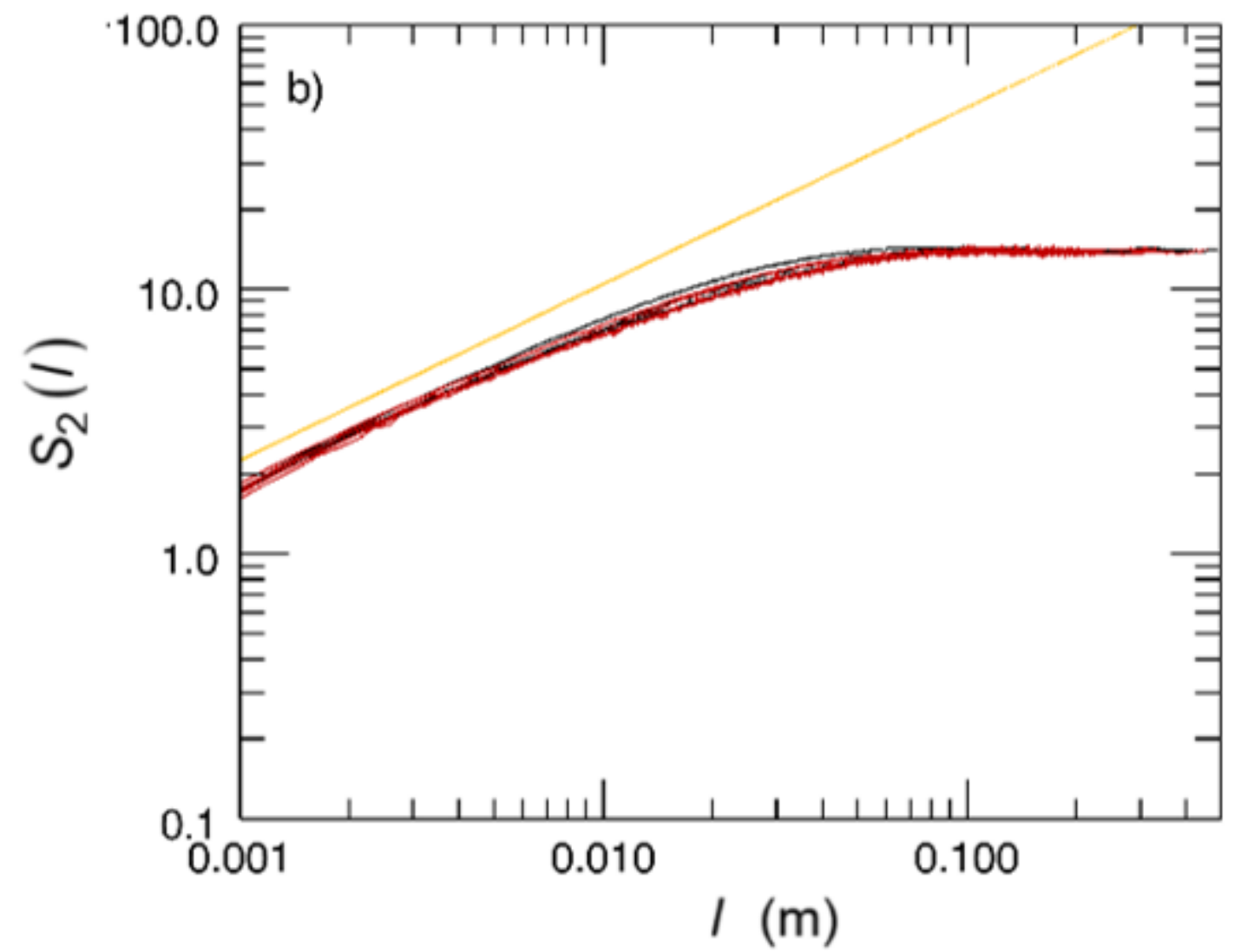}
    \end{minipage}
    \caption{a) Development of $S_2 (l)$ along the jet axis at distances $z/D=30 - 100$ (upper to lower). b) $S_2(l)$ scaled along abscissa and ordinate. Yellow line: $2/3$-slope.}
    \label{fig:16}
\end{figure}

Figure~\ref{fig:16}$a$ shows the development of the second order spatial structure function $S_2(l)$ along the centerline of the jet from $z/D = 30 : 10 : 100$. The $S_2(l)$-functions are observed to begin to level off at large separations, but they follow the expected $2/3$-slope predicted by the K41 theory for small separations $l$. Figure~\ref{fig:16}$b$ shows $S_2(l)$ scaled along the ordinate and the abscissa to convincingly collapse with $S_2(l)$ at $z/D=30$.

According to the simple model and the expected scaling of $S_2(l)$ from its dimension, $L^2T^{-2}$, we should expect the ordinate, $S_2$, to scale as the inverse second power of the local velocity scale, which we have taken to be the local mean velocity. The structures themselves, the abscissa $l$, we would expect to scale with the local spatial scale, which we have taken to be the local half width of the jet. 

\begin{figure}[t]
    \begin{minipage}{\linewidth}
    \centering
    \includegraphics[width=\linewidth]{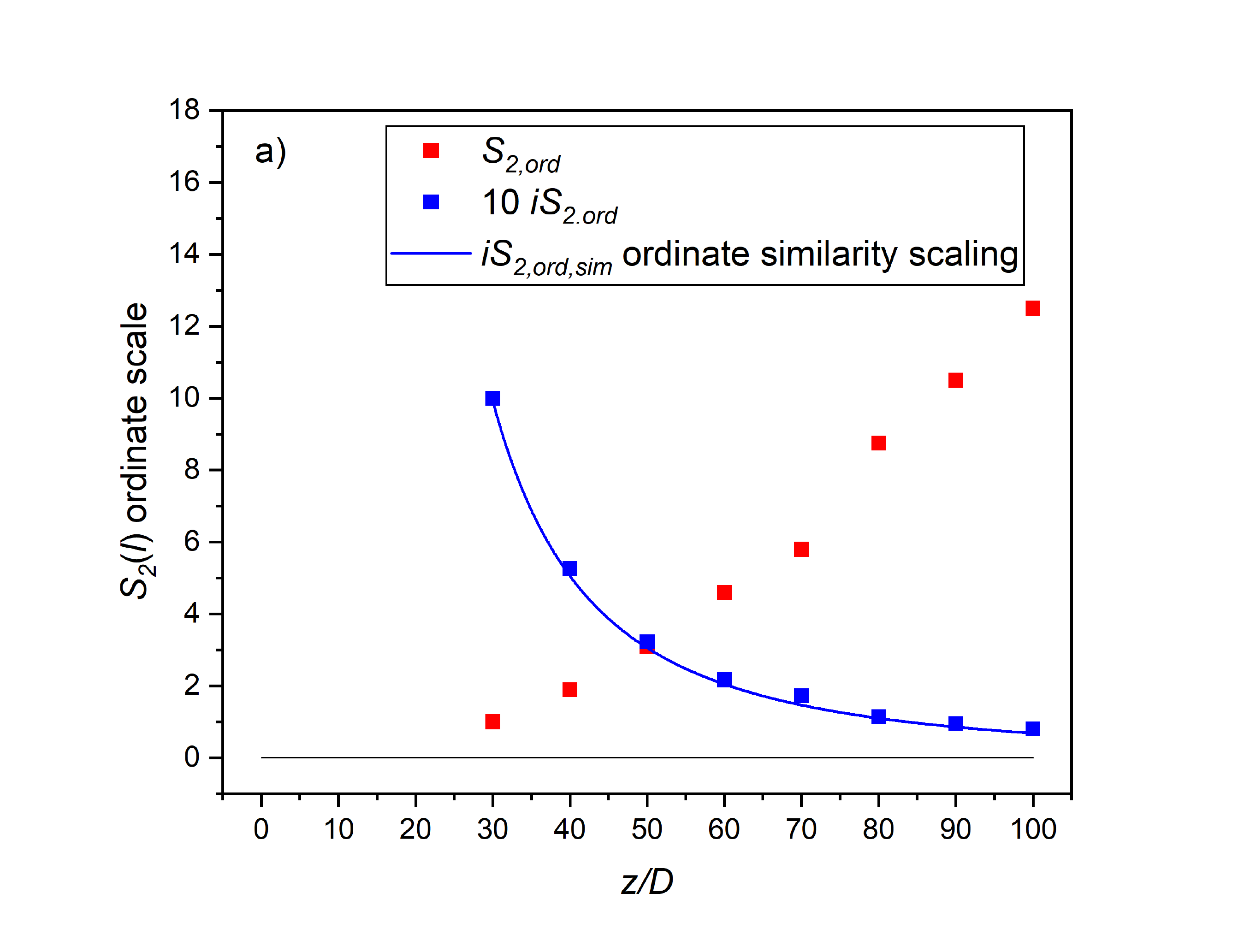}
    \end{minipage}
    \begin{minipage}{\linewidth}
    \centering
    \includegraphics[width=\linewidth]{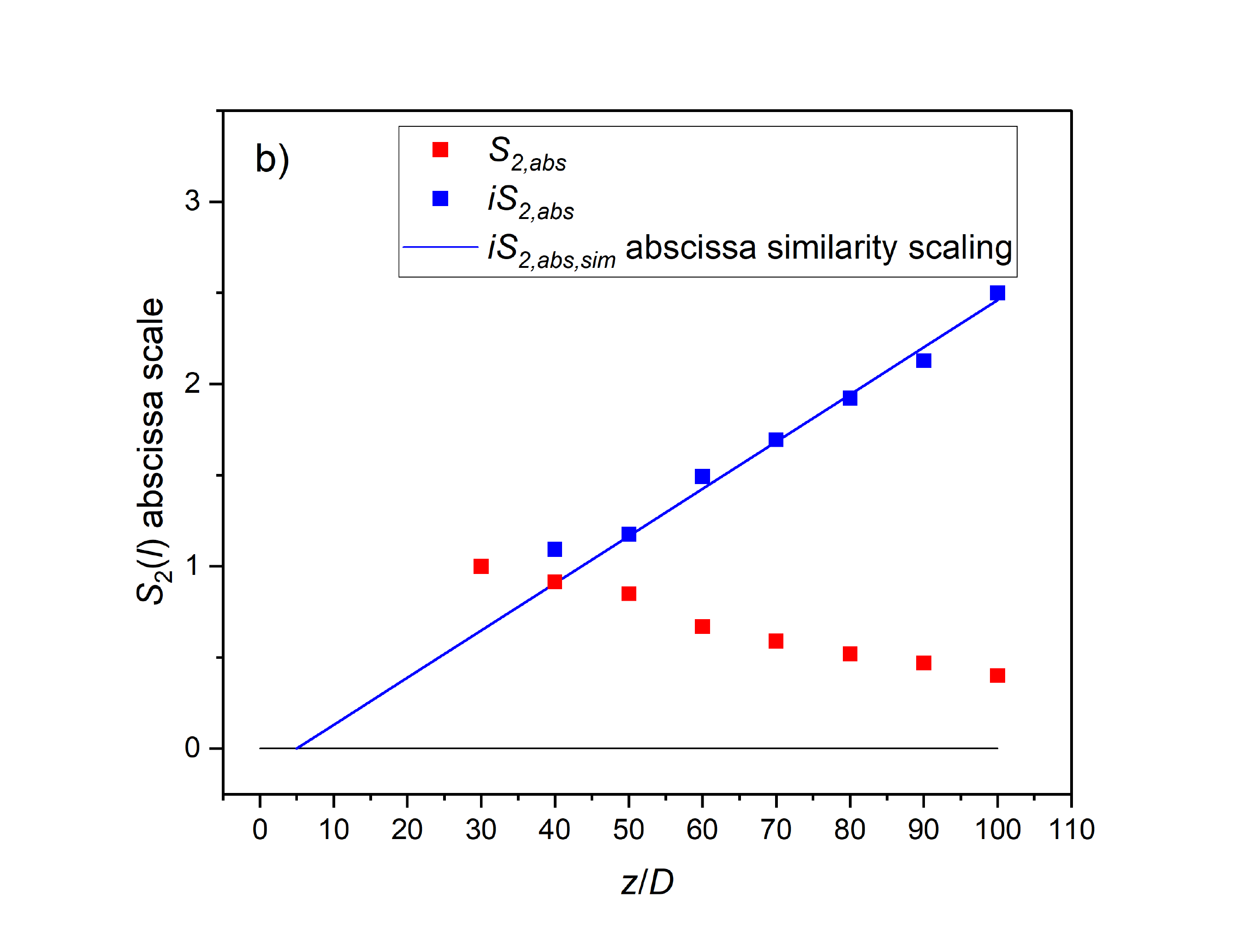}
    \end{minipage}
    \caption{a) $S_2(l)$ ordinate scaling factor, $S_{2,ord}$, (red squares) and the inverted ordinate scaling factor, $iS_{2,ord}$, multiplied by 10 (blue squares). The blue curve is the corresponding expected similarity scaling of the $S_2$ ordinate, $iS_{2,ord,sim}$. b) $S_2(l)$ abscissa scaling factor, $S_{2,abs}$, (red squares) and inverted abscissa scaling factor, $iS_{2,abs}$, (blue squares). Blue curve: Expected abscissa similarity scaling of spatial structures, $iS_{2,abs,sim}$.}
    \label{fig:17}
\end{figure}

In order to make the $S_2(l)$ plots collapse to the same ordinate value as that of $S_2(l)$ at $z/D = 30$, we have multiplied the ordinate of $S_2(l)$ at other positions along the jet axis with an ordinate factor, $S_{2,ord}$. Figure~\ref{fig:17}$a$ shows this scaling factor for all measured second order structure functions (red squares). The blue squares show the inverted scaling factor, $iS_{2,ord}$ (multiplied by 10). Also shown is the $S_2(l)$ scaling expected from the simple model,~\cite{JetSimPart1}, $iS_{2,ord,sim} = 6200\cdot (z-z_0)^{-2}$ with $z_0 / D = 5$ (blue curve in Figure~\ref{fig:17}$a$).  

Correspondingly, the factors needed to make $S_2(l)$ collapse along the abscissa with $S_2(l)$ at $z/D =30$, $S_{2,abs}$, is shown in Figure~\ref{fig:17}$b$ (red squares) along with the inverted scaling factor, $iS_{2,abs}$ (blue squares). The blue line in Figure~\ref{fig:17}$b$ shows the similarity scaling expected from the simple model,~\cite{JetSimPart1}, $iS_{2,abs,sim} = 0.0295 \cdot \left ( z-z_0 \right )$, with $z_0 / D =5$ (blue line in Figure~\ref{fig:17}$b$). The blue curve shows the growth of the spatial structures. The deviation at $z/D =30$ and $40$ can be attributed to the difficulty in visually distinguishing the very small shifts of the $S_2(l)$ plots at low displacement values. 

In conclusion, the scaling of $S_2(l)$ along both the abscissa direction and the ordinate direction shows excellent agreement with the simple geometrical one-factor scaling expected from the predictions of the simple jet model presented in~\cite{JetSimPart1}.

\begin{figure}[t]
    \begin{minipage}{\linewidth}
    \centering
    \includegraphics[width=\linewidth]{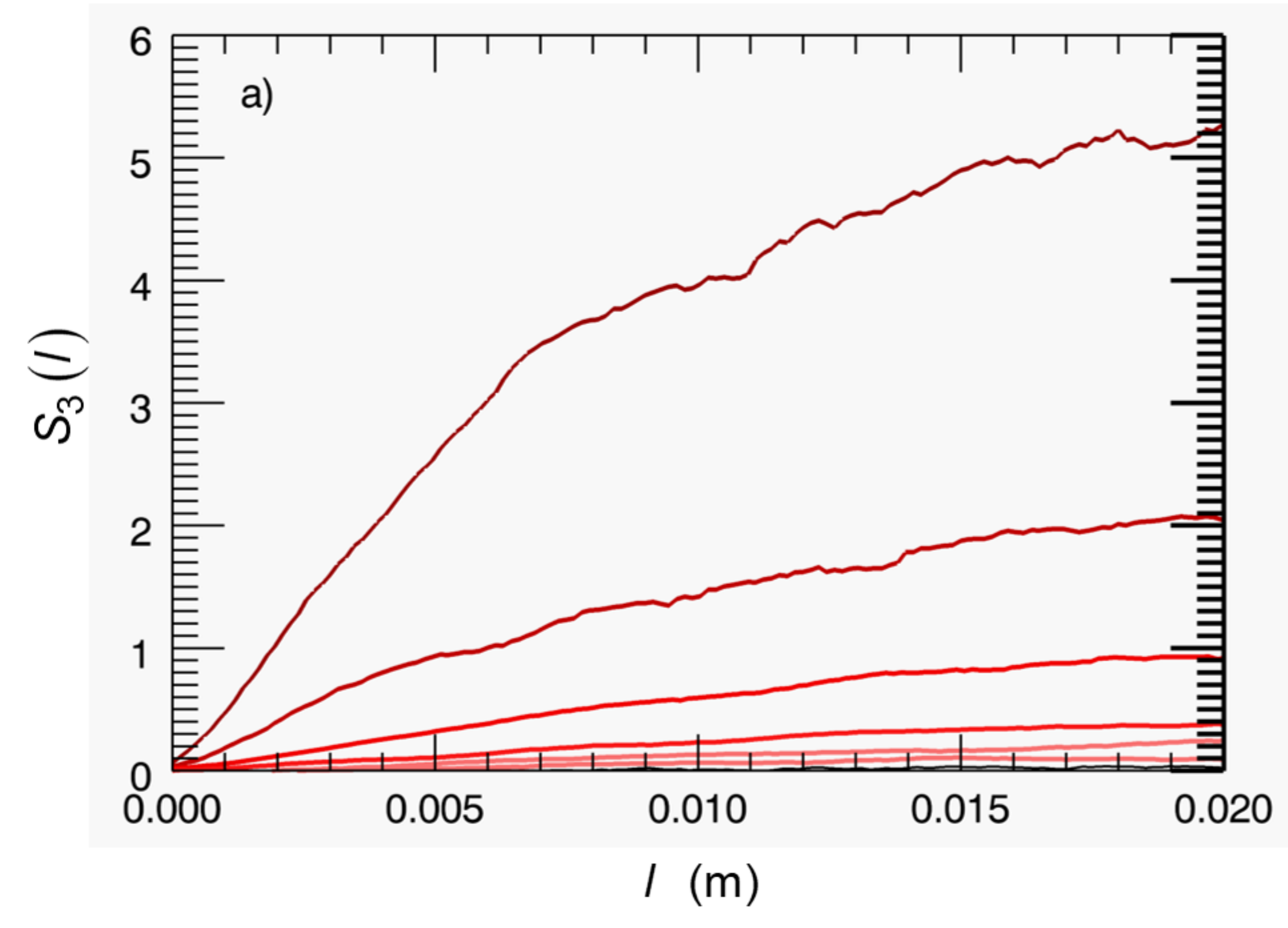}\vspace{0.3cm}
    \end{minipage}
    \begin{minipage}{\linewidth}
    \centering
    \includegraphics[width=\linewidth]{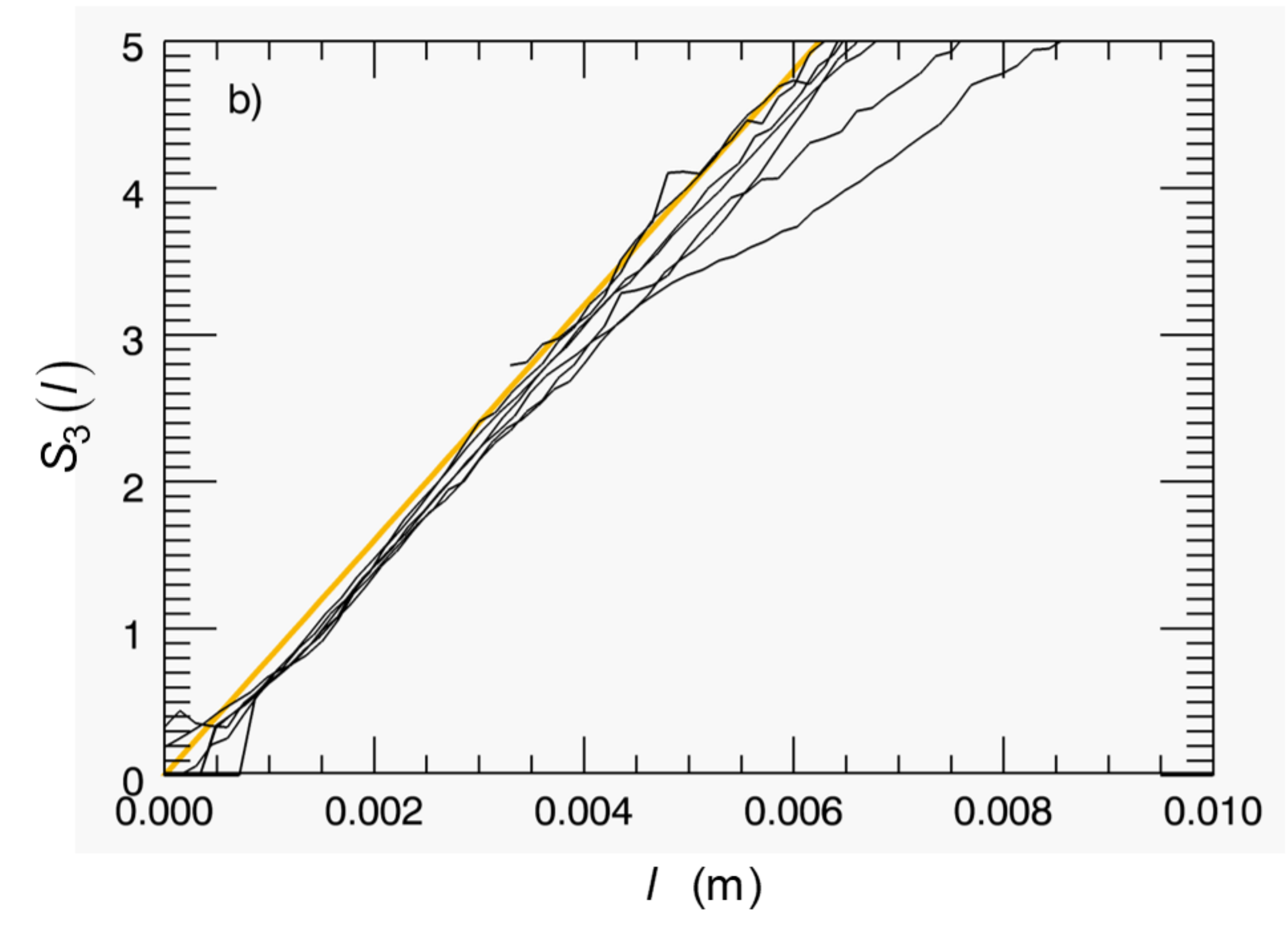}
    \end{minipage}
    \caption{a) Third order structure functions, $S_3(l)$, measured along the jet axis at distances $z/D=30 - 100$ (increasing downstream distance from upper to lower). b) Corresponding measured $S_3(l)$ scaled to have initial slope $4/5$ (yellow line). }
    \label{fig:18}
\end{figure}

Figure~\ref{fig:18} shows the third order structure function, $S_3(l)$, along the centerline of the jet as a function of the displacement, $l$. Figure~\ref{fig:18}$a$ shows the raw measured data. According to the K41 theory, for an equilibrium turbulent flow as is well approximated by the round jet, the third order structure function is expected to show a linear increase for small displacements with a slope proportional to the average rate of dissipation of turbulent kinetic energy, $\overline{\varepsilon}$: $S_3(l) = - \frac{4}{5}\overline{\varepsilon} l$. Figure~\ref{fig:18}$b$ shows a closeup of $S_3(l)$ for small lags rescaled to fit an initial slope of $4/5$ (yellow line): $\alpha S_3(l)/l=\frac{4}{5}$. Thus, the average dissipation rate is estimated as $\overline{\varepsilon}= 1/\alpha$. 

According to the dimension of $S_3(l)$, $L^3T^{-3}$, we can expect the $S_3(l)$-ordinate to scale as the inverse third power of the axial distance and for the spatial scales (the abscissa) to grow linearly with this distance. 

\begin{figure}[t]
    \begin{minipage}{\linewidth}
    \centering
    \includegraphics[width=\linewidth]{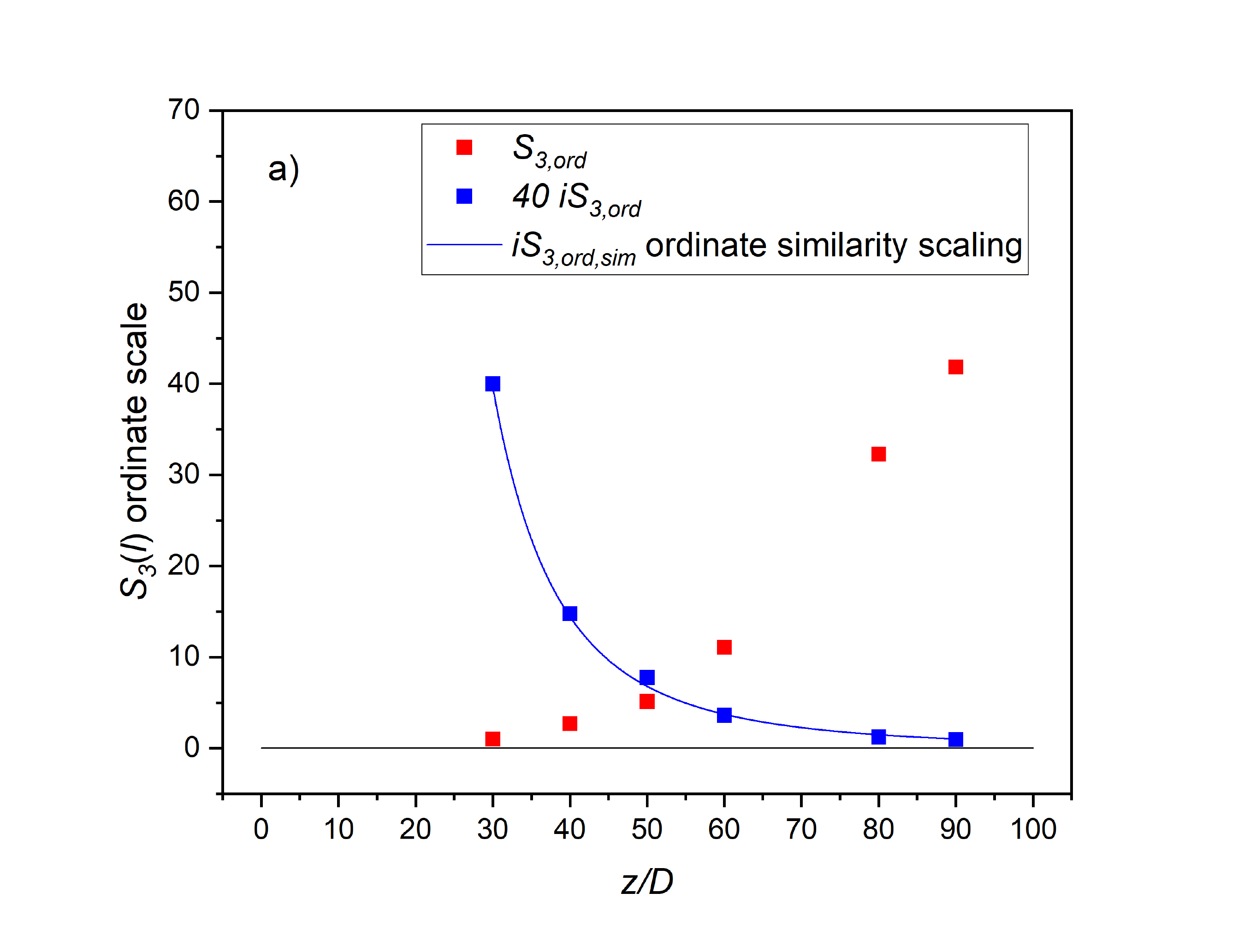}
    \end{minipage}
    \begin{minipage}{\linewidth}
    \centering
    \includegraphics[width=\linewidth]{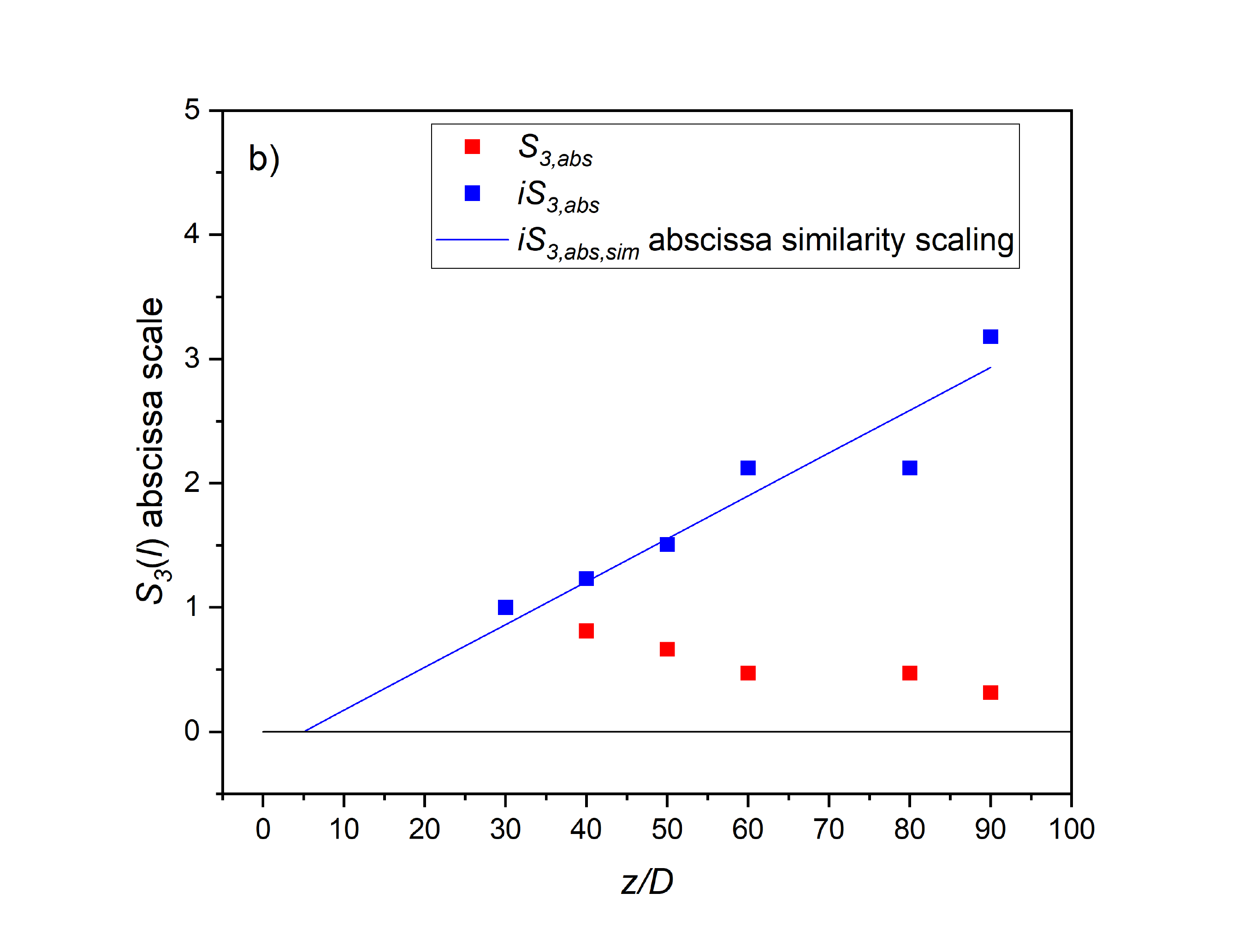}
    \end{minipage}
    \caption{a) Factors collapsing $S_3 (l)$ along the ordinate, $S_{3,ord}$ (red squares), the inverted $S_3$ scaling factor, $iS_{3,ord}$ (blue squares), and the expected ordinate similarity scaling from the simple model, $iS_{3,ord,sim}$ (blue line). b) The factors collapsing the abscissa, $S_{3,abs}$ (red squares), the inverted scaling factor, $iS_{3,abs}$ (blue squares), and the expected $S_3$ abscissa similarity scaling, $iS_{3,abs,sim}$ (blue line). Virtual origin $z_0/D=5$ in both cases.}
    \label{fig:19}
\end{figure}

Figure~\ref{fig:19}$a$ shows the factor, $S_{3,ord}$, needed to collapse the $S_3(l)$ ordinate to the $S_3(l)$-values at $z/D=30$, (red squares) and the inverse scaling factor, $iS_{3,ord}$, multiplied by $40$, (blue squares) and a similarity scaling function in agreement with the geometrical scaling expected from the simple model, $iS_{3,ord,sim} = 6.2\cdot 10^5 \cdot \left ( z-z_0 \right )^{-3} $ with $z_0/D=5$ (blue curve). 

Figure~\ref{fig:19}$b$ shows the factor, $S_{3,abs}$, needed to collapse $S_3(l)$ along the abscissa to the $S_3(l)$-values at $z/D =30$ (red squares), the inverse of this function, $iS_{3,abs}$ (blue squares), and a similarity scaling function, $iS_{3,abs,sim} = 0.034 \cdot \left ( z-z_0\right )$ (blue curve), again with $z_0/D=5$. The results of the $S_3(l)$-scaling show excellent agreement with a single geometrical scaling factor, i.e., the axial distance from a single virtual origin around five jet orifice diameters downstream. 

\begin{figure}[t]
    \centering
    \includegraphics[width=\linewidth]{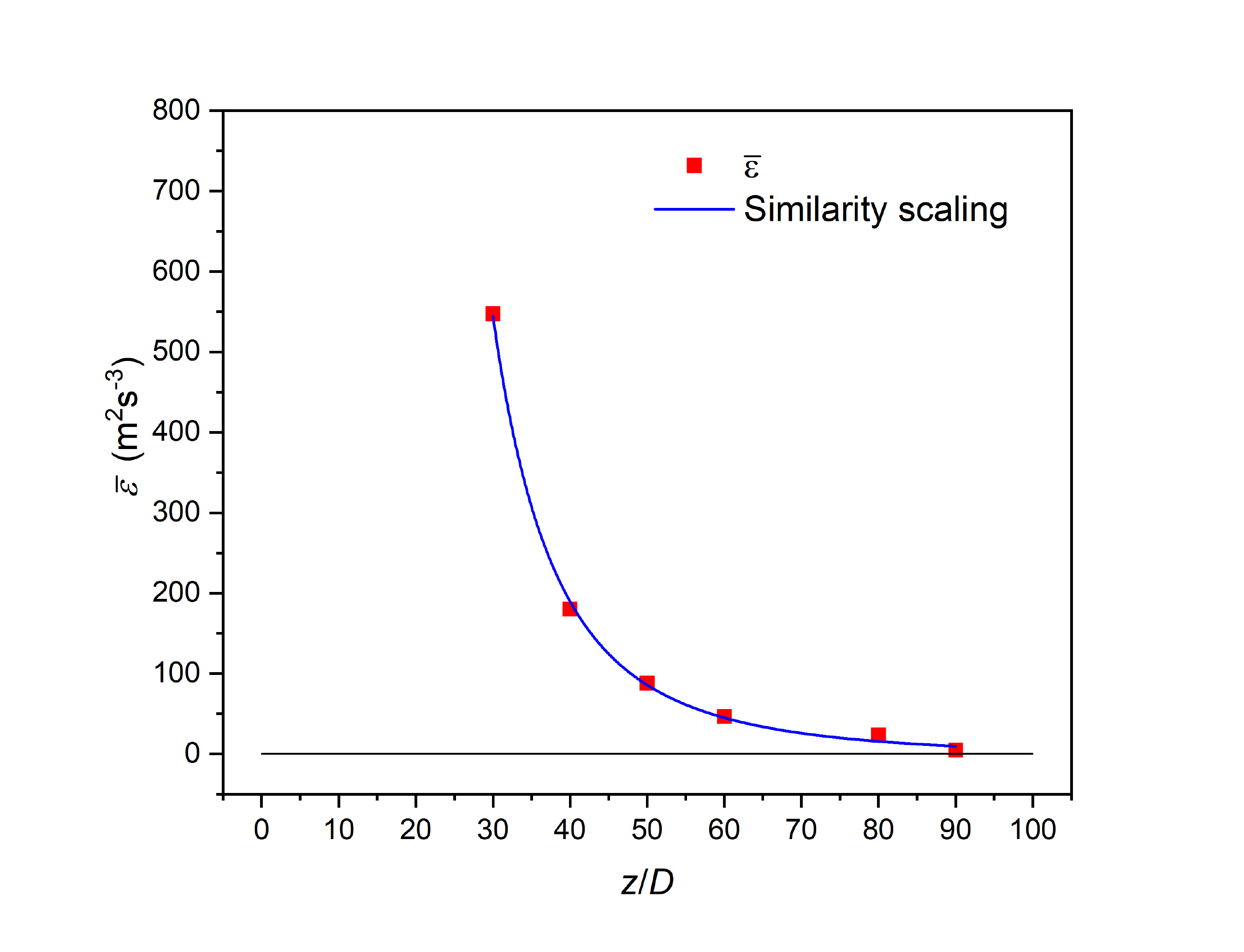}
    \caption{Downstream development of the average dissipation rate, $\overline{\varepsilon}$, along the centerline of the jet.}
    \label{fig:20}
\end{figure}

Figure~\ref{fig:20} shows the average dissipation rate, $\overline{\varepsilon}$, computed from the $S_3$-slope, $\alpha$. The data point at $z/D = 100$ was excluded due to the difficulty to fit the $S_3(l)$-slope accurately. 

\subsubsection{Spatial scales}

\begin{figure}[t]
    \centering
    \includegraphics[width=0.9\linewidth]{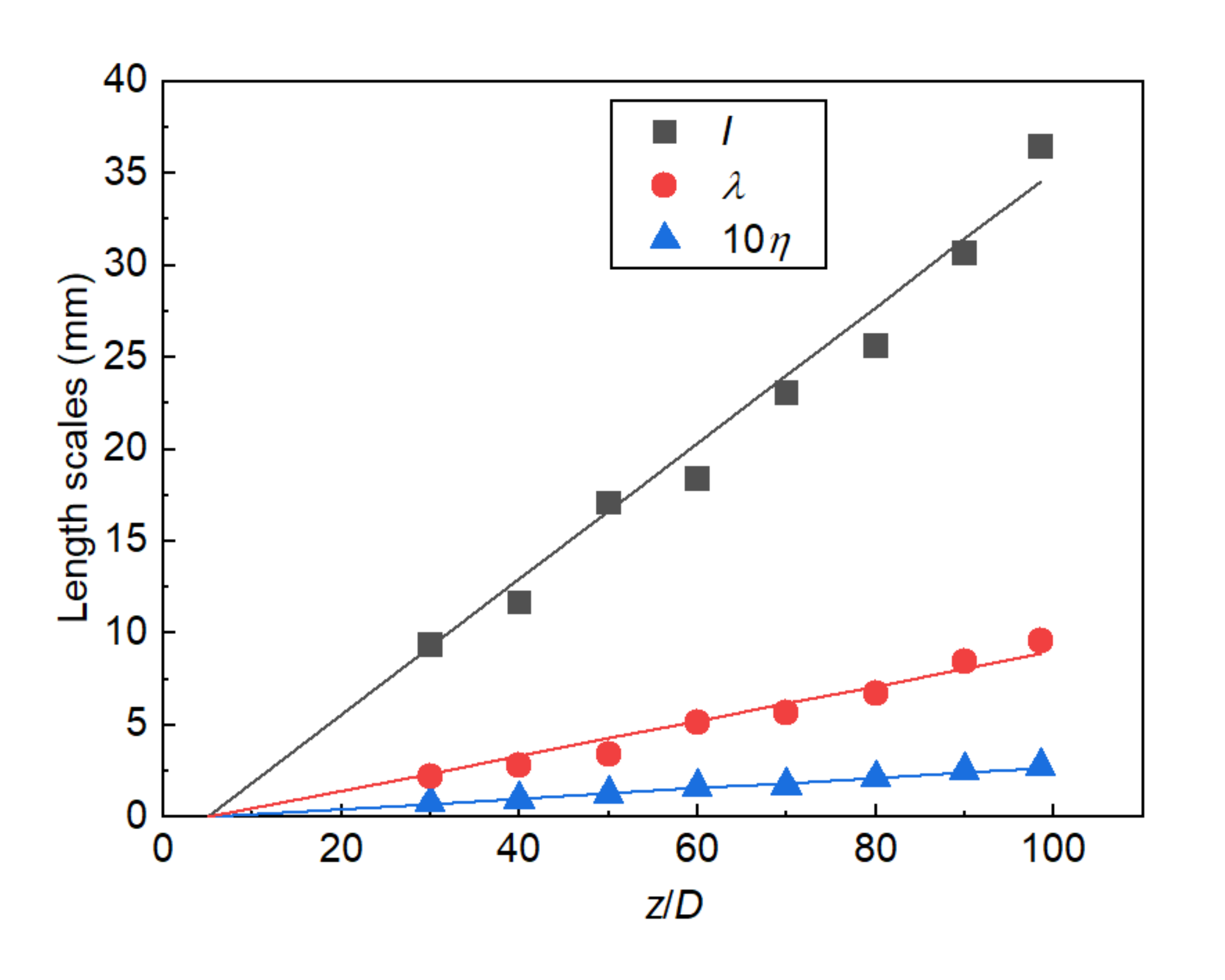}
    \caption{Downstream development of the integral scale, $I(z)$, the Taylor microscale, $\lambda (z)$, and the Kolmogorov microscale, $\eta (z)$ (the latter multiplied by a factor of 10 for readability). The downstream developments of the scales are compared to first order polynomials, all originating at $z_0 / D = 5$.}
    \label{fig:21}
\end{figure}

The measured time series in a stationary turbulent flow along with the convection record technique allow the computation of the spatial velocity scales as a function of the downstream distance. In all instances, the scales are computed for each convection record and afterwards averaged over all records. The downstream development along the centerline of the three length scales are displayed together in Figure~\ref{fig:21}; the integral scale, $I(z)$, the Taylor microscale, $\lambda (z) $ and the Kolmogorov microscale, $\eta (z)$.

The integral length scale is defined as the integral of the spatial autocorrelation function, $I \equiv \int_0^{\infty} C(l)\,dl$. We have used the ACFs, e.g. as shown in Figure~\ref{fig:15}, as basis for the calculation. The square of the Taylor microscale is computed in $k$-space, $\lambda^2 \equiv  \frac{\left \langle \overline{\hat{u}'(k)\hat{u}'(k)^{\ast}} \right \rangle}{\left \langle  \overline{k^2 \hat{u}'(k)\hat{u}'(k)^{\ast} } \right \rangle}$. The Kolmogorov microscale is computed from the third order structure function, $\eta = \left ( \frac{4 \nu^3}{5 | \alpha |} \right )^{1/4}$, where $\alpha$ is the initial slope of $S_3(l)$. Figure~\ref{fig:21} shows that all three scales have a linear evolution along the jet centerline with approximately the same virtual origin, $z_0 / D =5$, for all three scales, as expected from the model.

\section{\label{sec:Conclusion}Conclusions}
The purpose of the current work is to evaluate to what extent the statistical properties of a ``real'' jet in air correspond to predictions from a simple model governed by some basic symmetry properties and corresponding conservation laws~(\cite{JetSimPart1}). The predictions can be based on the dimension of the quantity as expressed by the order of velocity and length scales entering the quantity. Or stated differently, the abscissa and ordinate of the plots of the statistical functions shown in Table~\ref{tab:1}. The overall conclusion is that the scaling of all the quantities investigated in this work can be expressed by velocity, $u(z)$, and distance, $z$. Since the velocity is itself a function of $z$, the scaling of all the investigated statistical quantities can be said to depend on only one scaling parameter, namely the distance from the virtual origin, $z-z_0$. 

Careful laser Doppler anemometer measurements in a free, round jet in air have allowed collection of high-quality data, even in very high turbulence intensity and high shear regions of a high Reynolds number, turbulent jet. Conversion of the randomly sampled time records to spatial records, using the so-called convection record method, has allowed computation of single point static and dynamic first, second and third order spatially averaged statistical functions. Comparison of the measured statistical functions and the way they scale from one point in the jet to other points in the jet has convincingly shown that the jet, when isolated from external influence, appears to develop according to basic underlying symmetry properties governing Newtonian space and time and scale with a single geometrical scaling factor, the distance from a common virtual origin, $z_0$, which is to be determined from experiments.

\section*{Acknowledgements}
CMV acknowledges financial support from the European Research council: This project has received funding from the European Research Council (ERC) under the European Unions Horizon 2020 research and innovation program (grant agreement No 803419). 

CZ, YT and PB acknowledge financial support from the Poul Due Jensen Foundation: Financial support from the Poul Due Jensen Foundation (Grundfos Foundation) for this research is gratefully acknowledged.

\nocite{*}
\section*{References}
\bibliography{Main}

\end{document}